\documentclass[prd, twocolumn, nofootinbib, floatfix]{revtex4-1}

\usepackage[mathscr]{euscript}
\usepackage{amsmath}
\usepackage{graphicx}
\usepackage{dcolumn}
\usepackage{bm}
\usepackage{epsfig}
\usepackage{amssymb,latexsym,mathrsfs}
\usepackage{graphicx}
\usepackage{color}
\usepackage{hyperref}
\usepackage{float}
\usepackage{diagbox}
  
\definecolor{vividviolet}{rgb}{0.62, 0.0, 1.0}
\definecolor{amaranth}{rgb}{0.9, 0.17, 0.31}
\definecolor{palatinateblue}{rgb}{0.15, 0.23, 0.89}
\definecolor{brightpink}{rgb}{1.0, 0.0, 0.5}
\definecolor{cornflowerblue}{rgb}{0.39, 0.58, 0.93}
\definecolor{deepcarminepink}{rgb}{0.94, 0.19, 0.22}
\definecolor{radicalred}{rgb}{1.0, 0.21, 0.37}

\hypersetup{ linktoc=all,
    colorlinks, linkcolor={palatinateblue},
    citecolor={brightpink}, urlcolor={amaranth}
}

\newcommand{\be}{\begin{equation}}
\newcommand{\ee}{\end{equation}}
\newcommand{\bs}{\begin{split}} 
\newcommand{\bea}{\begin{eqnarray}}
\newcommand{\eea}{\end{eqnarray}}

\newcommand{\kp}{\kappa}

\renewcommand{\d}[1]{\ensuremath{\operatorname{d}\!{#1}}}

\begin{document}

\title{Light and Airy: a simple solution for relativistic quantum acceleration radiation}
\author{Michael R.R. Good${}^{1,2}$}
\author{Eric V.\ Linder${}^{2,3}$} 
\affiliation{${}^1$Physics Department, Nazarbayev University, Nur-Sultan, Kazakhstan\\
${}^2$Energetic Cosmos Laboratory, Nazarbayev University, Nur-Sultan, Kazakhstan\\ 
${}^3$Berkeley Center for Cosmological Physics \& Berkeley Lab, University of California, Berkeley, CA, USA
}

\begin{abstract} 
We study the quantum radiation of particle production by vacuum 
from an ultra-relativistic moving mirror (dynamical Casimir effect) solution 
that allows (possibly for the first time) analytically calculable time 
evolution of particle creation and an Airy particle 
spectral distribution. The reality of the beta Bogoliubov coefficients 
is responsible for the simplicity, and the mirror is asymptotically inertial 
at the speed of light, with finite energy production. We also discuss general relations 
regarding negative energy flux, the transformation to the 1-D 
Schr{\"o}dinger equation, and the incompleteness of entanglement entropy. 
\end{abstract} 

\date{\today} 

\maketitle 

\section{Introduction} 

Acceleration radiation with finite energy production is physically well-motivated.  In the case of black hole evaporation, for example, this is a conspicuous sign that the evolution has finished, energetic radiation has stopped, and conservation of energy is upheld. The canonical moving mirror model of DeWitt-Davies-Fulling \cite{DeWitt:1975ys, Davies:1976hi,Davies:1977yv}, for a single perfectly reflecting boundary point in flat (1+1)-D spacetime, has solutions demonstrating in a simple way total finite energy production (e.g.\ the four decade old solution of Walker-Davies which first derived a finite amount of energy creation \cite{Walker_1982}). Recently, several finite energy mirror solutions have been found that demonstrate close connections to strong gravitational systems.  These gravity analog models are called accelerated boundary correspondences (ABCs). The finite energy ABC solutions\footnote{The infinite energy ABC solutions correspond to the most well-known spacetimes, e.g.\ Schwarzschild \cite{Good:2016oey}, Reissner-Nordstr\"om (RN) \cite{good2020particle}, Kerr \cite{Good:2020fjz}, and de Sitter \cite{Good:2020byh}.} closely characterize interesting well-known curved spacetime end-states, including extremal black holes (asymptotic uniformly accelerated mirrors \cite{Liberati:2000sq,good2020extreme,Good:2020fjz,Rothman:2000mm,Foo:2020bmv}), black hole remnants (asymptotic constant-velocity mirrors \cite{Good:2016atu,Good:2018ell,Good:2018zmx,Myrzakul:2018bhy,Good:2015nja,Good:2016yht}) and complete black hole evaporation (asymptotic zero-velocity mirrors \cite{Walker_1982, Good:2019tnf,GoodMPLA,Good:2017kjr,good2013time,Good:2017ddq,Good:2018aer}).

Despite this progress, it has been very hard to find a mirror solution whose particle spectrum is simple.  Only two known solutions have analytic forms, one whose spectrum is an infinite sum of terms \cite{Good:2016yht} and another which is so lengthy as to be prohibitively cumbersome  \cite{Good:2019tnf,GoodMPLA}.  Consequently, analytic time evolution is impossible to find for the above spectra.  Further investigation of the particle production at any given moment is hobbled because one must instead resort to numerical analysis and finite sized frequency-time bins utilizing the discrete nature of orthonormal wave packets \cite{good2013time}.

Motivated by simplicity, we take a step back and consider that any Bogoliubov transformation can be broken down into two types: (1) the trivial unitary transformation with $\beta$ Bogoliubov coefficient zero, $\beta = 0$, indicating no particle production and (2) squeezing transformations where the $\beta \neq 0$ is given by a transformation matrix that is diagonal \cite{exact} (see the Bloch-Messiah decomposition or the theory of singular values). The simplest examples of the non-trivial transformations are those where the Bogoliubov coefficients are \textit{real-valued}. We therefore 
look for some mirror motion (i.e.\ ABC) that should lead to  
a real non-zero beta Bogoliubov coefficient for particle creation, 
and anticipate corresponding simplicity in the resulting spectrum.

We take the simplest possible choice for global 
mirror motion with characteristics leading to the 
desired reality of the Bogoliubov coefficient, and 
indeed find a simple solution for the particle 
production spectrum. 
Remarkably, a transformation to the time domain on this spectrum analytically gives the particle production at any given moment.  

Our paper is organized as follows: in 
Sec.~\ref{sec:preview} we give a very brief motivation 
of the connection between the reality of the beta Bogoliubov 
coefficient  and the mirror trajectory 
properties. We analyze this accelerated trajectory in 
Sec.~\ref{sec:motion}, computing the key relativistic dynamical properties such as rapidity, speed, and acceleration. In Sec.~\ref{sec:energy} we derive the energy radiated, by analysis of the quantum stress tensor, and in Sec.~\ref{sec:particles} we derive the particle spectrum, finding a unique Airy-Ai form for the radiation and confirming consistency with the stress tensor results. Finally, in Sec.~\ref{sec:time} we compute the time evolution of particle creation analytically. Appendices~\ref{sec:apxsum} and 
\ref{sec:apxzero} discuss some general properties 
leading to necessary negative energy flux, and 
connecting to the 1-D Schr{\"o}dinger equation, 
respectively. Appendix~\ref{sec:apxentropy} is a note on the connection between rapidity and entanglement entropy.  Throughout we use natural units, $\hbar = c = 1$.

\section{Reality, Acceleration, and Inertia} \label{sec:preview}

The beta Bogoliubov coefficient controls quantum particle 
production. In light-cone coordinates $(u,v)$, with retarded 
time $u=t-x$ and advanced time $v=t+x$, the moving mirror 
trajectory $f(v)$ gives retarded time position, and the beta 
Bogoliubov coefficient is \cite{Birrell:1982ix} 
\be 
\beta_{\omega\omega'} = \frac{-1}{4\pi\sqrt{\omega\omega'}}\int_{-\infty}^{+\infty} \d v ~e^{-i \omega' v -i \omega f(v)}\left(\omega f'(v)-\omega'\right)\,,\label{betaint} 
\ee 
where $\omega$ and $\omega'$ are the frequencies of the outgoing and incoming modes respectively \cite{carlitz1987reflections}.

To maintain finite energy and the simplicity of no information loss, 
there must not be a horizon at finite time, and the acceleration 
must vanish at infinity (i.e.\ the mirror motion must be 
asymptotically inertial). Under these conditions we can carry out 
an integration by parts to give 
\be 
\beta_{\omega\omega'} = \frac{1}{2\pi}\sqrt{\frac{\omega'}{\omega}}\int_{-\infty}^{+\infty} \d v\: e^{-i\omega'v-i\omega f(v)}\,.\label{partsint} 
\ee  

To guarantee a real-valued beta Bogoliubov coefficient, the mirror trajectory $f(v)$ must be an odd function so that the exponential over the symmetric interval turns into a cosine of the argument, i.e.\ a real valued function. 
The simplest odd function that accelerates in the required manner 
is $f(v) \sim v + v^3$. We will find this results in not only 
interesting dynamics, but analytic calculation of particle 
production spectrum and time evolution.

\section{Trajectory Motion}\label{sec:motion} 

As motivated in the previous section, we expect the accelerated 
mirror trajectory  
\be 
f(v) = v + \kappa^2\frac{ v^3}{3}\,,\label{f(v)} 
\ee  
to have interesting physical properties. Here $\kp$ is a 
quantity related to the acceleration (and the surface gravity 
in the black hole case). 

We can also 
write the trajectory in spacetime coordinates, 
\be 
t=-x+\frac{1}{\kp}(-6\kp x)^{1/3}\,, 
\ee 
taking the real cube root, or 
\be 
x=-t-\frac{1}{2\kp}\left[A_+^{2/3}A_-^{1/3}+A_+^{1/3}A_-^{2/3}\right]\,, \label{x(t)} 
\ee 
where 
\be 
A_\pm=3\kp t \pm\sqrt{9\kp^2 t^2+8}\,. 
\ee 
Note at late times $x\to -t+{\mathcal O}(t^{1/3})$. 
These forms make it obvious that asymptotically the mirror travels at the speed of light. 

\begin{figure}[!th]
\centering
\includegraphics[width=\columnwidth]{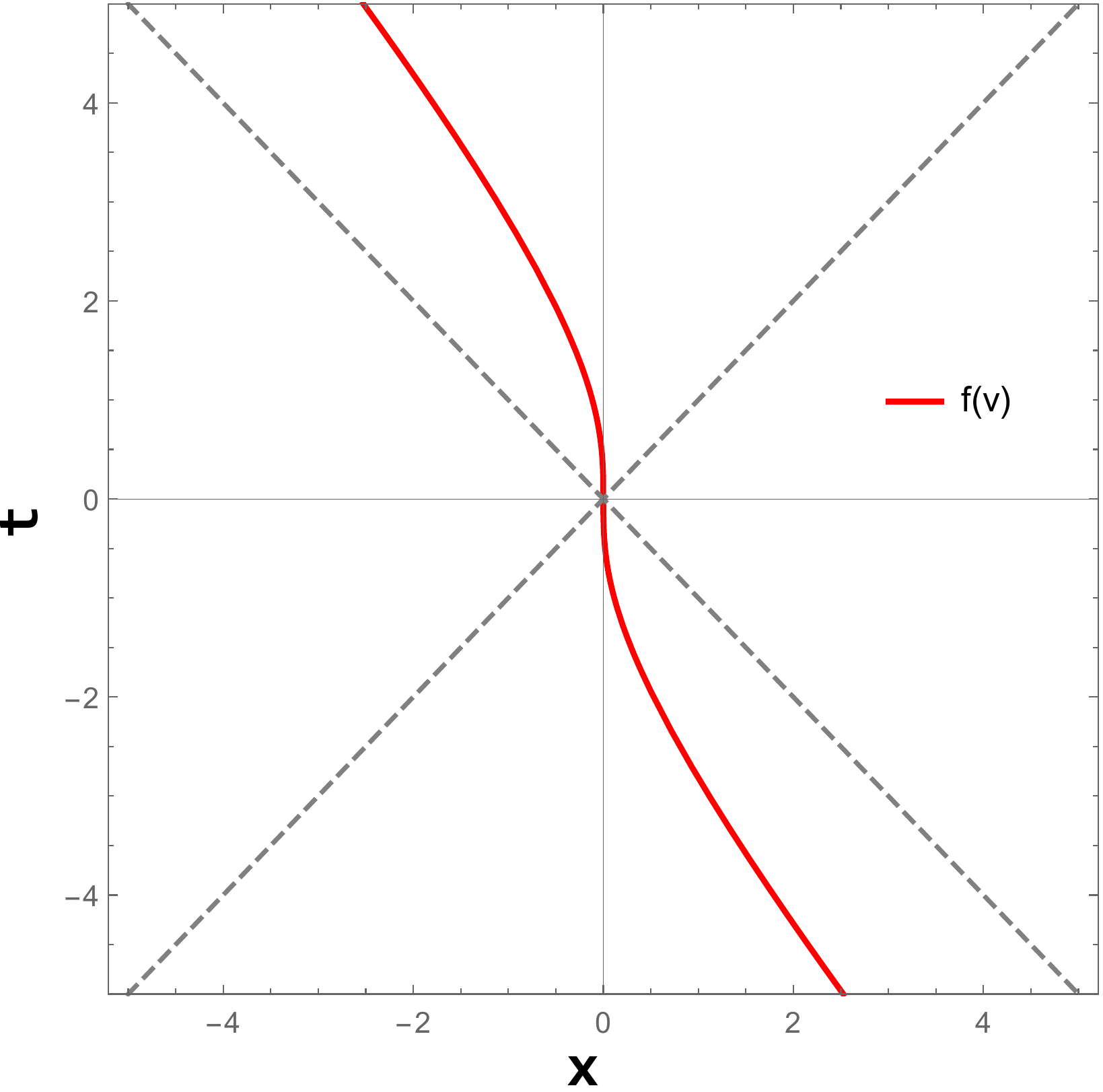}
\caption{A spacetime diagram of the mirror trajectory, Eq.~(\ref{f(v)}) with $\kappa=1$. 
It starts off asymptotically inertial with zero acceleration and light-speed velocity and decelerates, eventually reaching zero speed (at $t=0$), and then accelerates again approaching the speed of light in an asymptotically inertial way.  Note that field modes moving to the left will always hit the mirror, demonstrating no horizon, despite the mirror accelerating to light-speed. 
}\label{Fig1}
\end{figure}
\begin{figure}[!ht]
\centering
\includegraphics[width=\columnwidth]{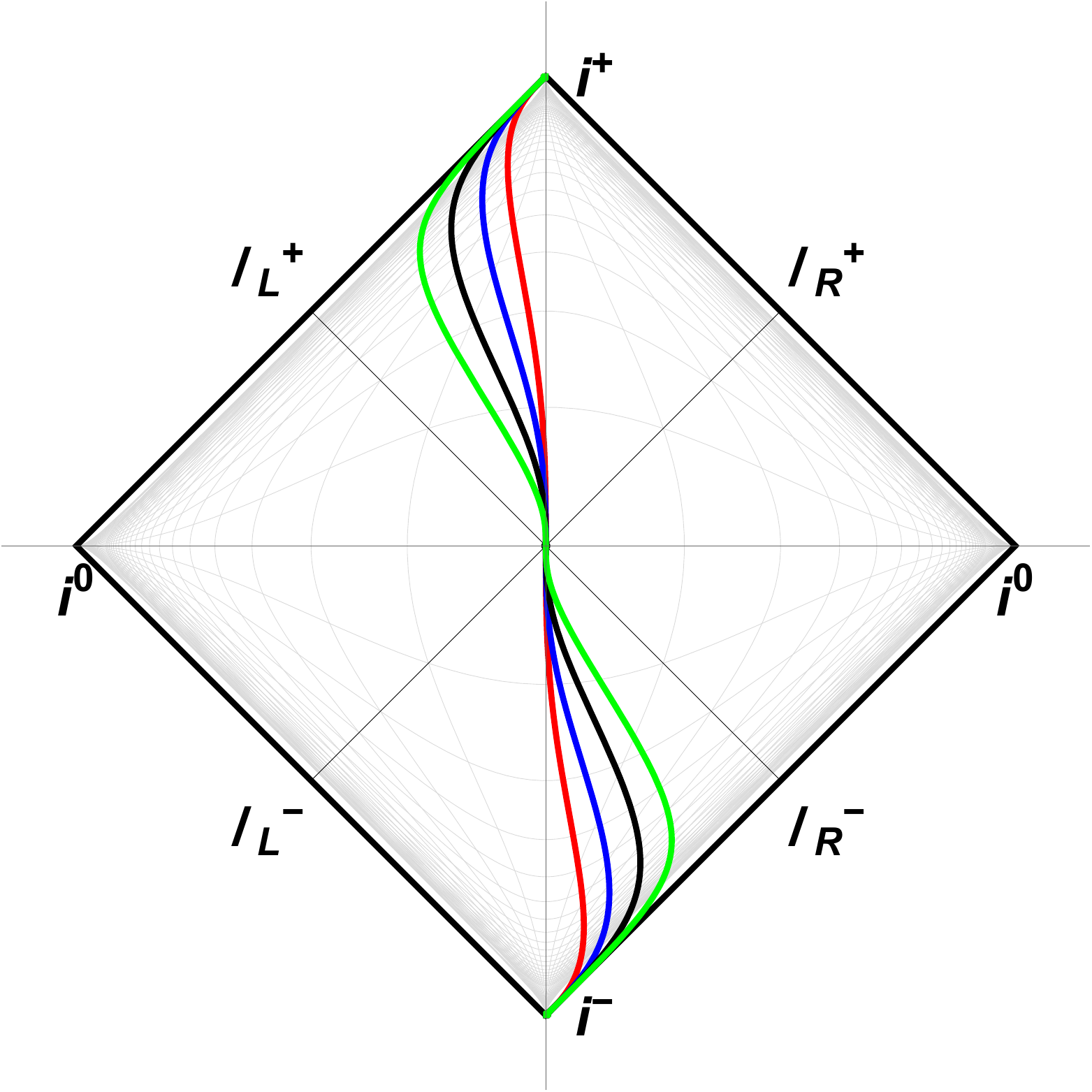}
\caption{A Penrose diagram of the mirror trajectory, Eq.~(\ref{f(v)}). The mirror is moving at light-speed at $v\to\pm\infty$. 
Since the acceleration is asymptotically zero as $v\to \pm\infty$ then this mirror is  asymptotically inertial.  The various colors correspond to different maximum accelerations; here $\kappa = 1,4,16, 64$ from red, blue, black, and green. 
}\label{Fig2}
\end{figure}

A spacetime plot with time on the vertical axis is given of the trajectory in Figure \ref{Fig1}. A conformal diagram is plotted in Figure \ref{Fig2}.  We next investigate the dynamics of the trajectory Eq.~(\ref{f(v)}). 

We compute the rapidity $\eta(v)$ by $2\eta(v) \equiv  -\ln f'(v)$ where the prime is a derivative with respect to the argument, 
\be \eta(v) = -\frac{1}{2} \ln \left(\kappa ^2 v^2+1\right).\label{eta(v)}\ee
From the rapidity we may easily compute the velocity $V \equiv \tanh \eta$, plugging in Eq.~(\ref{eta(v)}), 
\be    
V(v) = -\tanh \left[\frac{1}{2}\ln \left(\kappa ^2 v^2+1\right)\right] = 
\frac{-\kp^2 v^2}{2+\kp^2 v^2}\,, \label{V(v)} 
\ee  
and the proper acceleration, which follows from $\alpha(v)\equiv  e^{\eta(v)} \eta'(v)$, 
\be \alpha(v)=-\frac{\kappa ^2 v}{\left(\kappa ^2 v^2+1\right)^{3/2}}\,.\label{alpha(v)}
\ee 

\begin{figure}[!ht]
\centering
\includegraphics[width=\columnwidth]{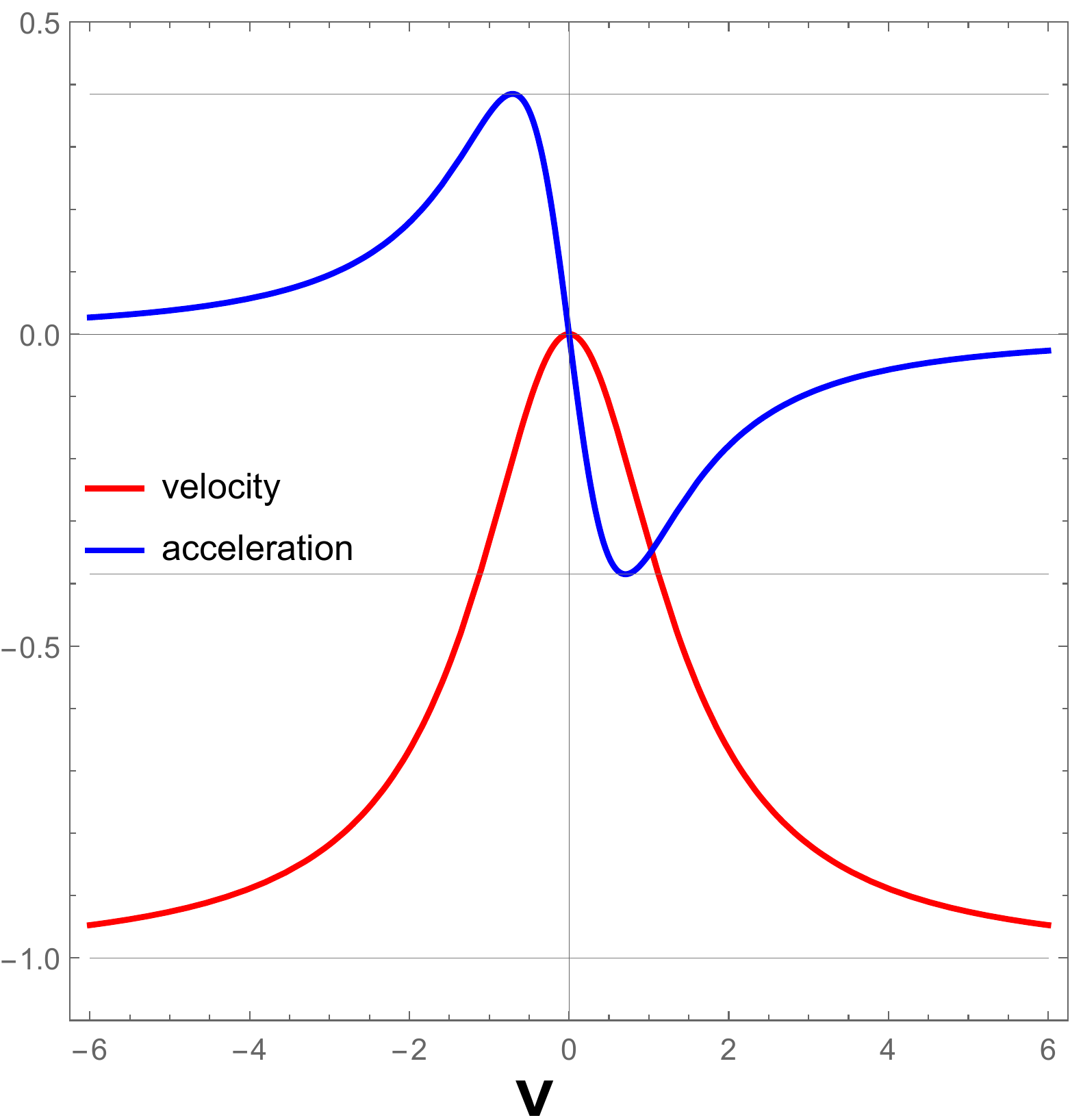}
\caption{The velocity and proper acceleration as a function of light-cone coordinate advanced time $v=t+x$ for the mirror trajectory, Eq.~(\ref{f(v)}).  At $v=0$, the velocity $V=0$, but asymptotically $|V|\to 1$ and the proper acceleration vanishes, $\alpha \to 0$. The maximum acceleration occurs at $|\alpha_{\textrm{max}}| = 2\kappa/(3\sqrt{3}) = 0.385\kappa$.  Here 
$v$ is in units of $1/\kappa$ and the maximum accelerations happen at advanced time $ \kp v = \pm 1/\sqrt{2} = 0.707$. 
}\label{Fig3}
\end{figure}

At $x=t=0=v$, the velocity and acceleration are zero. 
At asymptotic infinity, the velocity is the speed of light and 
the acceleration goes to zero. 
The magnitude of the velocity, Eq.~(\ref{V(v)}), along with the proper acceleration, Eq.~(\ref{alpha(v)}), are plotted in Figure~\ref{Fig3}.

\section{Energy Flux and Total Energy}\label{sec:energy} 

\subsection{Energy Flux} 

\begin{figure}[H]
\centering
\includegraphics[width=\columnwidth]{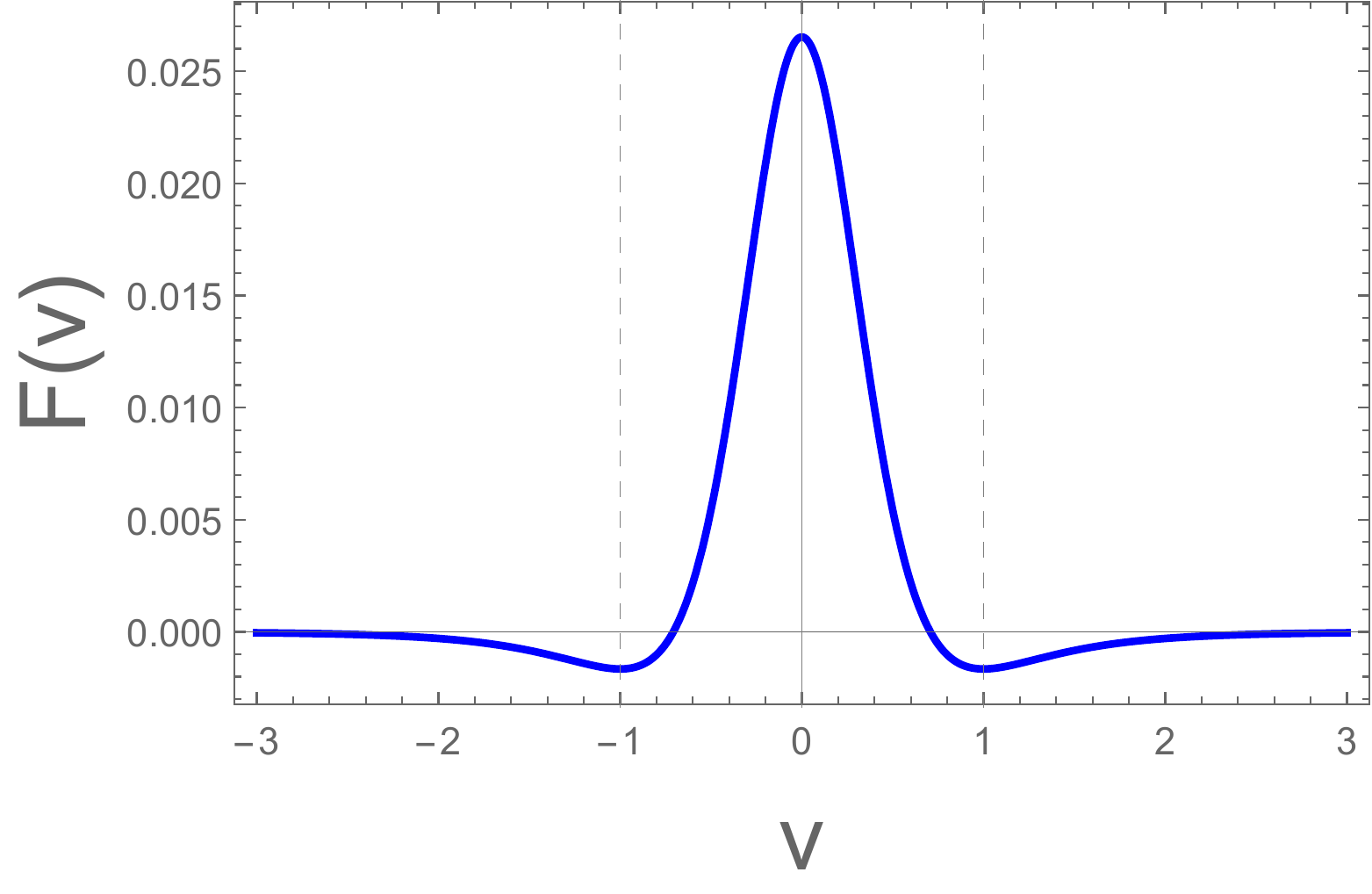}
\caption{The energy flux, Eq.~(\ref{F(v)exact}), is asymptotically zero at $v=\pm \infty$.  The total energy, as we shall see in Eq.~(\ref{totalE1}), is therefore finite, $E=\kappa/96$.  Notice the emission of negative energy flux near early and late advanced times. The maximum flux $F_\textrm{max} = \kappa^2/(12\pi)$ occurs at $v=0$ and the minimum flux $F_\textrm{min} = -\kappa^2/(192\pi)$ occurs at $v=\pm 1/\kp$. The energy flux crosses zero, $F=0$, at $v=\pm 1/(\sqrt{2}\kappa)$. Here $\kappa=1$.}
\label{Fig4}
\end{figure}

The quantum stress tensor reveals the energy flux emitted by the moving mirror.  Typically, one will see \cite{Davies:1976hi}
\be F(u) = -\frac{1}{24\pi}\{p(u),u\}, \label{F(u)}  
\ee 
where the energy flux, $F(u)$, is a function of light-cone coordinate retarded time $u = t-x$ \cite{Davies:1977yv, Birrell:1982ix} and the brackets define the Schwarzian derivative.  The trajectory in light-cone coordinates of the mirror is $p(u)$ which is the advanced time position ``$v$" as a function of retarded time $u$.  However, since we want advanced time $v$ as the independent variable, we write the radiated energy flux using $f(v)$  \cite{Good:2016atu,Good:2020byh}, 
\be F(v)= \frac{1}{24\pi}\{f(v),v\}f'(v)^{-2}\,,\label{F(v)}\ee
where the Schwarzian brackets are defined as usual,
\be \{f(v),v\}\equiv \frac{f'''}{f'} - \frac{3}{2}\left(\frac{f''}{f'}\right)^2\,.\ee 
For $f(v)$ given by Eq.~(\ref{f(v)}), this yields  
\be F(v) = \frac{\kp^2}{12\pi}\,\frac{1-2 \kappa ^2 v^2}{\left(\kappa ^2 v^2+1\right)^4}\ .\label{F(v)exact}\ee
It is clear that asymptotically $F(v)\to 0$ for both $v\to \pm \infty$. 
Figure~\ref{Fig4} shows 
the energy flux as a function of advanced time $v$.

\subsection{Total Energy} 

The total energy measured by a far away observer at $\mathscr{I}^+_R$ is \cite{walker1985particle} 
\be E = \int_{-\infty}^{\infty} F(u) \d u\,, \ee
where integration occurs over retarded time (it takes the energy time to reach $\mathscr{I}^+_R$).  Since we are using advanced time $v$, we write this with $du = \frac{\d f}{\d v} dv$ to get the Jacobian correct, 
\be E = \int_{-\infty}^{+\infty} F(v) f'(v) dv\,.\label{totalE}\ee
Plugging in Eq.~(\ref{f(v)}) and Eq.~(\ref{F(v)exact}) into Eq.~(\ref{totalE}), with Jacobian $du/dv = \kappa ^2 v^2+1$, the simple result is 
\be E = \frac{\kappa}{96}\,,\label{totalE1}\ee
which is finite and positive. 

Physically, the finite value tells us the evaporation process stops, similar to the ABC's of extremal black holes (asymptotic uniformly accelerated mirrors), black hole remnants (non-horizon sub-light-speed asymptotic coasting mirrors), and complete black hole evaporation (asymptotic static moving mirrors). The fact that the total energy is positive is consistent with the quantum interest conjecture \cite{Ford:1999qv} as derived from quantum inequalities \cite{Ford:1994bj}.

\subsection{Negative Energy Flux} 

As seen from Figure \ref{Fig4}, there are regions of negative energy flux (NEF). 
This is required by the unitarity sum 
rule (see Appendices~\ref{sec:apxsum} and \ref{sec:apxzero} and, e.g., \cite{Good:2019tnf}). 
These regions extend for $|v|>1/(\kp\sqrt{2})$.  
The total negative energy is, by symmetry, 
\be E_{NEF} = 2\int_{v=+\frac{1}{\kp\sqrt{2}}}^{v=+\infty} F(v) f'(v) \d v\,,\label{NEFtotal}\ee
which gives an analytic result 
\be E_{NEF} =\frac{\kappa  \left(-10 \sqrt{2}+3 \pi -6 \cot ^{-1}\sqrt{2}\right)}{288 \pi } = -0.00930 \kappa \ee 
As a ratio, the emission of NEF to positive energy flux (PEF) is 
\be \frac{|E_{NEF}|}{E_{PEF}} \approx 47.1\%\,.\ee 
Note one cannot judge by eye this ratio in Figure~\ref{Fig4} 
due to the redshift Jacobian $f'(v)$ in Eq.~(\ref{NEFtotal}). 

Another interesting aspect is that because the rapidity 
diverges, so does the entropy flux $S=-\eta/6$. However, 
since there is no horizon there is no information loss. 
This indicates that entanglement entropy is not a 
comprehensive measure of the unitary, finite energy, 
information preserving dynamics, due to the inertial light 
speed asymptote (see Appendix~\ref{sec:apxentropy}).

\section{Particle Spectrum}\label{sec:particles} 

The particle spectrum can be obtained from the beta Bogoliubov coefficient, given by Eq.~\eqref{partsint} in 
Sec.~\ref{sec:preview}. 
For the particular trajectory Eq.~\eqref{f(v)}, as promised the Bogoliubov coefficient is real, 
\be 
\beta_{\omega\omega'} = \frac{-1}{(\omega\kp^2)^{1/3}}\sqrt{\frac{\omega'}{\omega}}\, 
\text{Ai}\left(\frac{\omega +\omega '}{(\omega\kp^2)^{1/3}}\right)\,, 
\label{betaai} 
\ee 
which is highly unusual. This corresponds to the Bogoliubov 
transformation being a pure boost without rotation, i.e. there is no phase on the beta coefficient, giving us a natural choice for both field modes and coefficients (and potentially an action integral whose real part defines the vacuum–vacuum amplitude \cite{Nikishov:2002ez}). 

\begin{figure}[H]
\centering
\includegraphics[width=\columnwidth]{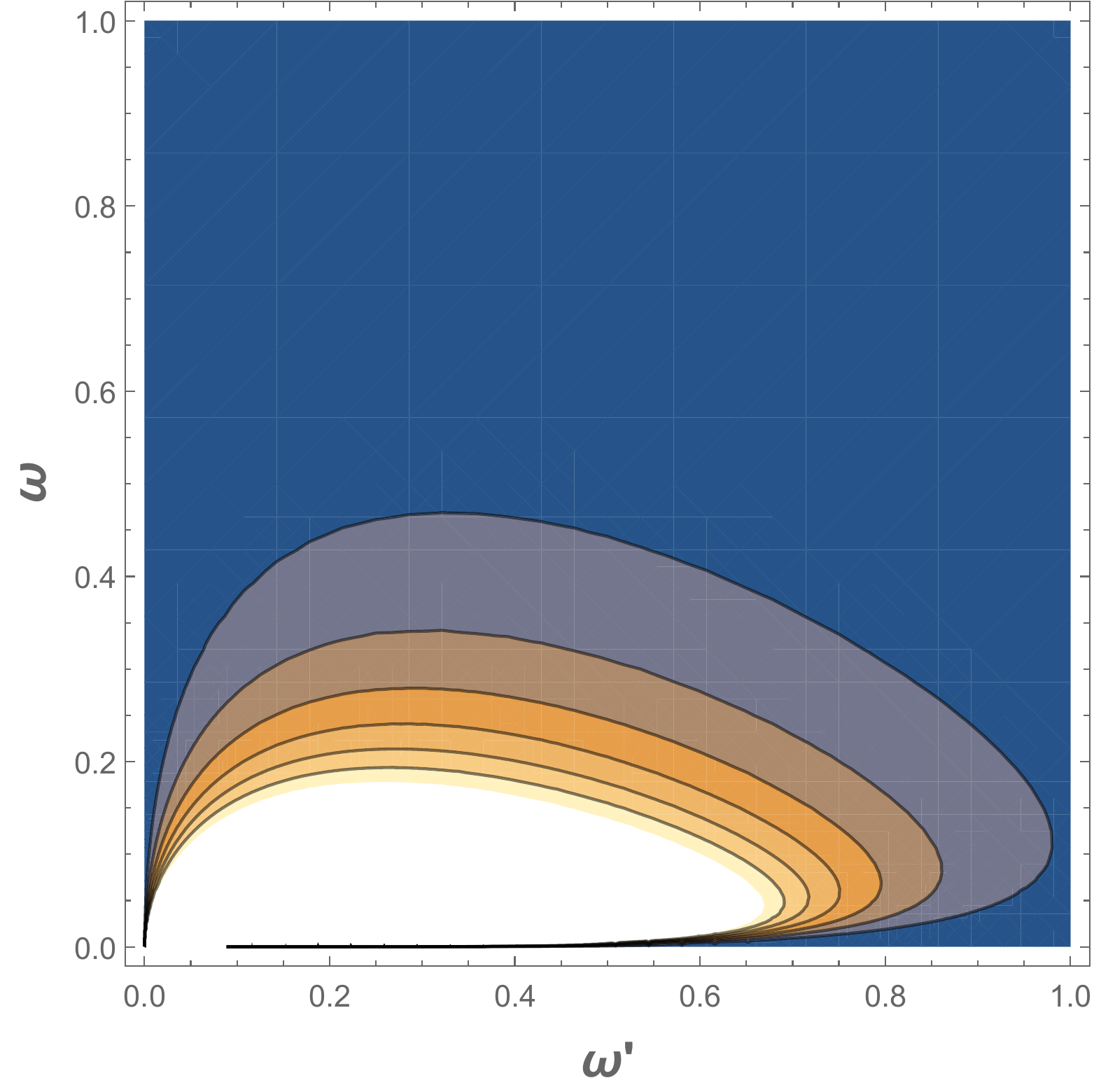}
\caption{The Airy-Ai spectrum,  $|\beta_{\omega\omega'}|^2$ from 
Eq.~(\ref{spectrum}), 
as a contour plot, here with $\kappa=1$.  
The brighter the contours the more particle production.  Notice the asymmetry between $\omega$ and $\omega'$ which are uniformly scaled.  This asymmetry ultimately shows up in the infinite total particle count due to the infrared divergence of $\omega$ in $N_\omega$ but makes it possible to analytically integrate $N_{\omega\omega'}$ over $\omega'$.
}\label{Fig5}
\end{figure}

To obtain the particle spectrum, we take the modulus square, $N_{\omega \omega'} \equiv |\beta_{\omega\omega'}|^2$, which gives 
\be  N_{\omega \omega'} = \frac{\omega '}{\kappa ^{4/3} \omega ^{5/3}}\,  \text{Ai}^2\left(\frac{\omega +\omega '}{\kappa ^{2/3} \omega^{1/3 }}\right).\label{spectrum}\ee
The Airy-Ai function is perhaps most well-known as the solution to the time-independent Schr{\"o}dinger equation for a particle confined within a triangular potential well and for a particle in a one-dimensional constant force field.\footnote{The triangular potential well solution is directly relevant for the understanding of electrons trapped in semiconductor heterojunctions.} 
The spectrum Eq.~(\ref{spectrum}), $|\beta_{\omega\omega'}|^2$, is explicitly non-thermal and plotted as a contour plot in Figure~\ref{Fig5}.

This demonstrates a new spectrum of radiation emanating from a moving mirror trajectory.  Eq.~(\ref{spectrum}) can be compared to the late time (equilibrium after formation) spectra of non-extremal black holes (e.g. Schwarzschild, RN, Kerr),
\be N_{\omega\omega'} = \frac{1}{2\pi \kappa \omega'}\frac{1}{e^{2\pi \omega/\kappa}-1}\,,\ee
and extremal black holes (e.g. ERN, EK, EKN),
\be N_{\omega\omega'} =\frac{e^{-\pi\omega c/\mathcal{A}}}{{\pi^2\mathcal{A}^2}}\ \left|K_{1+i\omega  c/\mathcal{A}}\left(\frac{2}{\mathcal{A}}\sqrt{\omega\omega'} \right)\right|^2\,.
\ee 
(For EK, $c=\sqrt{2}$; for ERN, $c=2$; for EKN, $c=\mathcal{A}/\kp$.)  Here $\kappa$ is the surface gravity, i.e.\  $\kappa = 1/(4M)$ in the case of a Schwarzschild  black hole, or outer horizon surface gravity for the RN and Kerr non-extremal black holes. 
In addition, $\mathcal{A}$ is the extremal parameter, or the asymptotic uniform acceleration \cite{Foo:2020bmv} in the case of the mirror system, while $K_\nu$ is the modified Bessel function of the second kind with order $\nu$. 

\begin{figure}[H]
\centering
\includegraphics[width=\columnwidth]{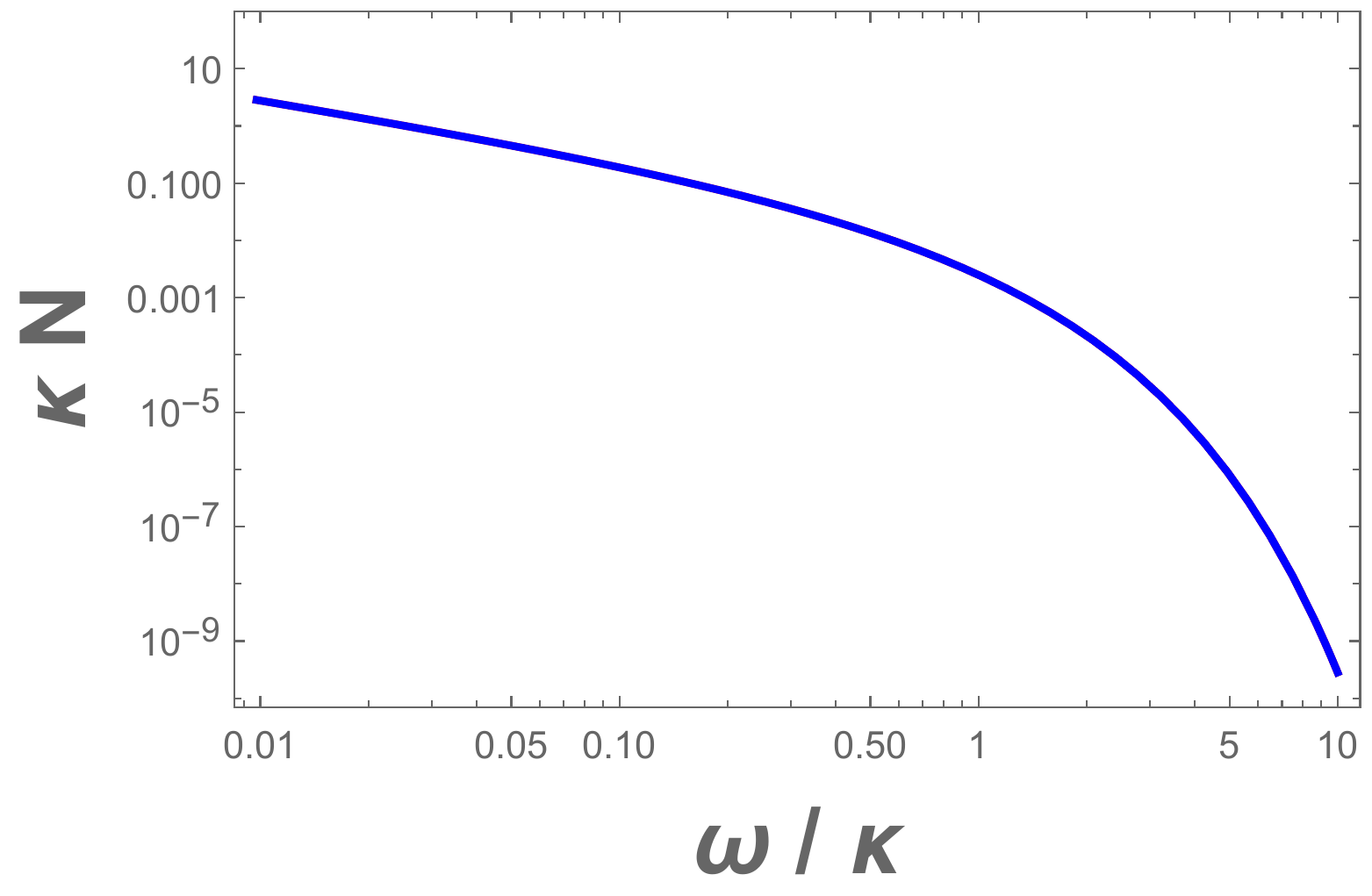}
\caption{The Airy particle spectrum, $N(\omega)$, Eq.~(\ref{Nw}). 
Note the infrared divergence at $\omega \to 0$; 
the soft particle divergence results in infinite total particle count characteristic of asymptotic coasting mirrors.  Larger maximum acceleration as measured by $\kappa$ results in more particles for a wider range of frequencies, i.e. $N$ scales as $1/\kp$ as seen by the $\kp N$ curve plotted vs $\omega/\kp$. }\label{Fig6}
\end{figure}

Furthermore, it is remarkable that the spectrum
\be N_\omega = \int_0^{\infty} N_{\omega\omega'} d\omega'\,,\ee
is analytic,  
\be N_\omega = \frac{2\sqrt{\bar{\omega}}}{3\kappa} \text{Ai}^2(\bar{\omega})-\frac{\text{Ai}(\bar{\omega}) \text{Ai}'(\bar{\omega})}{3 \kappa \bar{\omega}^{3/2}}-\frac{2 \text{Ai}'^2(\bar{\omega})}{3\kappa \sqrt{\bar{\omega}}} \,, \label{Nw}\ee
where $\bar{\omega} \equiv (\omega/\kappa)^{2/3}$. This analytic $N(\omega)$ spectrum is plotted in Figure~\ref{Fig6} for all $\kappa$.

The Airy functions can be reformulated into Bessel functions using the identities 
\bea 
{\rm Ai}(x)&=&\sqrt{\frac{x}{3\pi^2}}\,K_{1/3}\left(\frac{2}{3}x^{3/2}\right)\\ 
{\rm Ai}'(x)&=&\frac{-x}{\sqrt{3\pi^2}}\,K_{2/3}\left(\frac{2}{3}x^{3/2}\right)\,. 
\eea 
This turns Eq.~\eqref{spectrum} into 
\be 
N_{\omega\omega'}=\frac{1}{3\pi^2\kp^2}\frac{q'(q+q')}{q^2}\,K^2_{1/3}\left(\frac{2(q+q')^{3/2}}{3q^{1/2}}\right)\,, 
\ee 
which has similarities to the extremal black hole expression. 
Here $q=\omega/\kp$, $q'=\omega'/\kp$. For the particle spectrum we get 
\bea  
9\pi^2\kp\,N_\omega&=&2qK^2_{1/3}(2q/3)+K_{1/3}(2q/3)\,K_{2/3}(2q/3)\notag\\ 
&\qquad&-2qK^2_{2/3}(2q/3)\,. \label{eq:nw} 
\eea  

In the small and large $\omega$ limits the leading order terms are, respectively, 
\bea 
N_\omega &\to& \frac{1}{6\sqrt{3}\pi\omega}\,,\qquad \omega\to 0\,,\\ 
&\to& \frac{\kappa  }{16 \pi  \omega^2}\,e^{-4 \omega/(3 \kappa )}\,,\qquad \omega\to\infty\,. 
\eea 
The $1/\omega$ in the small frequency limit  
(note this is independent of $\kp$) demonstrates the infrared divergence leading to an infinite total particle count commonly associated with constant-velocity moving mirror solutions \cite{Good:2016atu,Good:2018ell,Good:2018zmx,Myrzakul:2018bhy,Good:2015nja,Good:2016yht}, that are not asymptotically static (asymptotic zero-velocity  \cite{Walker_1982, Good:2019tnf,GoodMPLA,Good:2017kjr,good2013time,Good:2017ddq,Good:2018aer}).  

To check that the energy is indeed carried away by the particles, we look for consistency  between 
Eq.~(\ref{spectrum}) and 
the total energy, Eq.~(\ref{totalE1}), found from the stress tensor. This is done by quantum summing,
\be E = \int_{0}^{\infty}\int_{0}^{\infty} \omega N_{\omega\omega'} \d \omega \d \omega',\label{qsum}\ee
that is, associating a quantum of energy $\omega$ with the particle distribution and integrating over all the frequencies. The result is pleasingly analytic:
\be 
E = \frac{\kappa}{96}\,.\label{Nw_E} 
\ee 
Since this is also the result of Eq.~(\ref{totalE1}), the beta spectrum Eq.~(\ref{spectrum}), or 
Eq.~\eqref{eq:nw}, is consistent with the quantum stress tensor, Eq.~(\ref{F(v)exact}). 

\begin{figure}[ht]
\centering 
\includegraphics[width=\columnwidth]{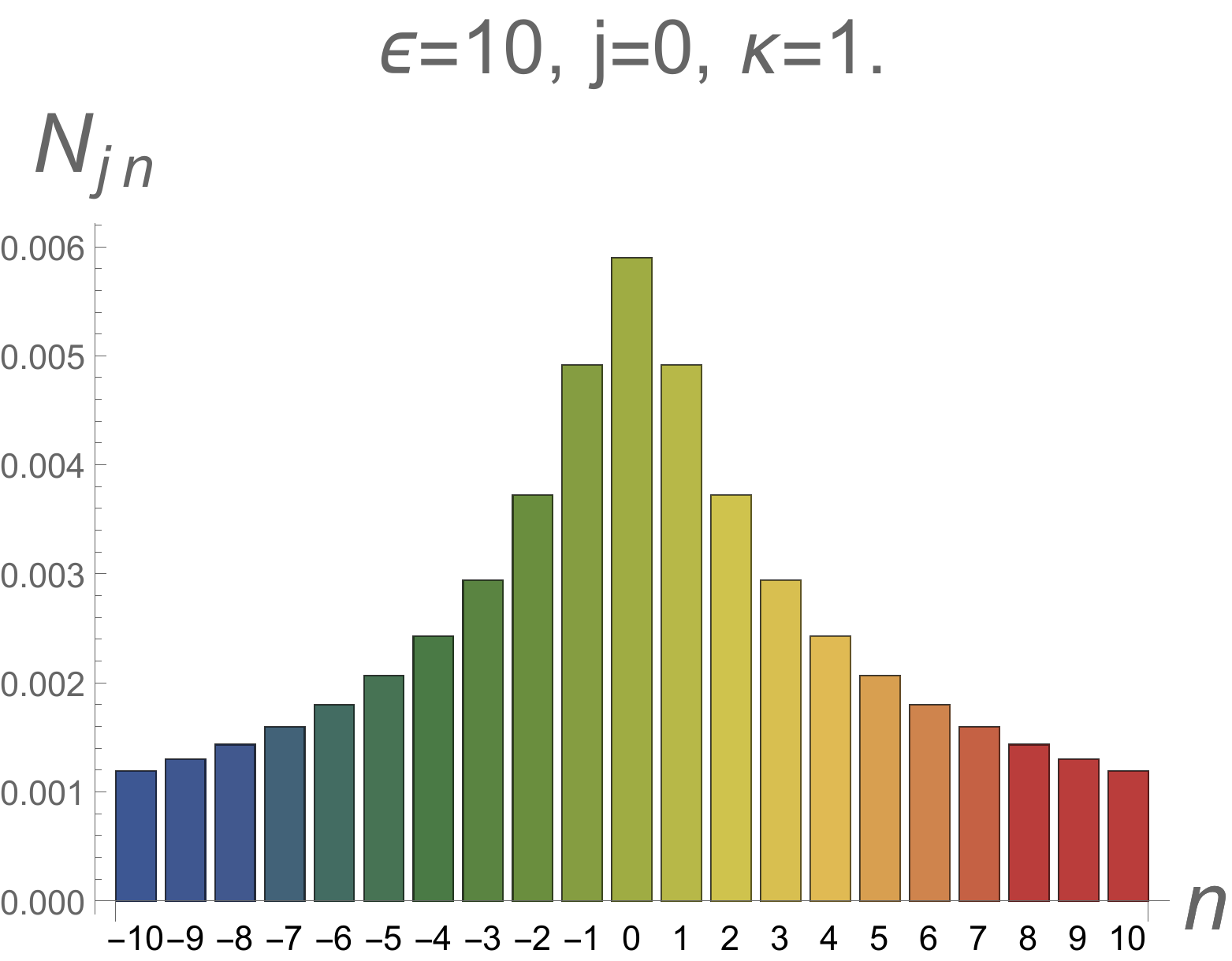} \caption{ The particle count in time, via wave packet localization.  The detector is set with $\epsilon = 10$, a relative large value ($\epsilon > 1$) in order to get clear time resolution. The scale of the system is $\kappa =1$ and the frequency bin is in the lowest possible $j=0$ value, where most of the particle production occurs, and finer resolution in time is possible. Notice there is no plateau, hence indicative of non-thermal radiation.  This emission includes the `phantom radiation' of soft particles as described in \cite{Liberati:2000sq}.  It is symmetric in delayed time, $u$, centered around time bin $n=0$.    
}
\label{Fig7} 
\end{figure}

The time dependence of particle creation can be computed via wavepacket analysis treated in Hawking \cite{Hawking:1974sw}, and explicitly numerically computed in \cite{GoodMPLA, Good:2019tnf}. Wave packet localization, particularly via orthonormal and complete sets in the moving mirror model, was first carried out in detail in \cite{Good:2012cp}.  For completeness, we utilize the same code to illustrate particle creation in time and present the results in Figure~\ref{Fig7}.  The rate of emission of particles is finite only in a given time and frequency interval which can be seen by these complete orthonormal family of wave packets constructed from the beta Bogoliubov coefficients, following Hawking's notation,
\be \beta_{jn \omega'} = \frac{1}{\sqrt{\epsilon}}\int_{j\epsilon}^{(j+1)\epsilon} d\omega\, e^{2\pi i \omega n/\epsilon} \beta_{\omega\omega'}\,,\label{betapacket}\ee
where $j \geq 0$ and $n$ are integers. These packets are built at future right null infinity, $\mathscr{I}^+_R$, and peak at delayed exterior time, $u = 2\pi n/\epsilon$, with width $2\pi/\epsilon$.  Therefore the vertical axis in Figure~\ref{Fig7} has a discrete and intuitive physical interpretation, giving the counts of a particle detector sensitive to only frequencies within $\epsilon$ of $\omega_j = j\epsilon$, for a time $2\pi/\epsilon$ at $u = 2\pi n/\epsilon$. Late times correspond to large quantum number $n$ 
(for the mirror Eq.~(\ref{x(t)}), late times have $u\approx 2t[1+{\mathcal O}(\kp t)^{-2/3}]$).  
For excellent time resolution, only one frequency bin is needed, where the particles pile up, $j=0$, and a relatively large value of $\epsilon$ resolves the count in time.  The text of Fabbri-Navarro-Salas \cite{Fabbri} also describes the details needed to reconstruct Figure~\ref{Fig7} by first packetizing the beta coefficient as done in Eq.~(\ref{betapacket}) and then secondly numerically integrating over $\omega'$ from $0$ to $\infty$, and third, computing the results, $N_{jn}$,
\be N_{j n} = \int_0^{+\infty}d\omega' |\beta_{jn \omega'}|^2,\label{Njn}\ee
for each individual time bin, $n$, for a set frequency bin, $j$ (in our fine-grained time resolution case, $j=0$). While this numerical approach evolves the particle count in time, it is not particularly stream-lined, fast, nor arbitrarily accurate.  In Sec.~\ref{sec:time} we will find an analytic approach to the evolution process, resolving these issues.

\section{Analytic Time Evolution}\label{sec:time} 

The spectrum, Eq.~(\ref{Nw}), is simple enough that analytical time evolution without discrete wave packetization is possible -- 
possibly uniquely in the literature.  
Typically we would like to employ a Fourier transform converting from frequency to time.  Since this does not work out in a  straightforward manner, we consider that the Fourier transform of a radially symmetric function in the plane can be expressed as a Hankel transform. The Hankel transform, $N_u =H(N_\omega)/2$ -- 
where by time symmetry we have divided the spectrum by 2 so that retarded time $u$ ranges from $-\infty$ to $+\infty$ -- 
is analytically tractable for the spectrum Eq.~(\ref{Nw}):
\bea
\frac{384}{\kappa}N_u &=&\nonumber5 \, _3F_2\left(\frac{7}{6},\frac{3}{2},\frac{11}{6};1,2;-\frac{9}{16} u^2 \kappa ^2\right)\\ \nonumber
&+& 4 \, _3F_2\left(\frac{1}{2},\frac{5}{6},\frac{7}{6};1,1;-\frac{9}{16} u^2 \kappa ^2\right)\\
&-&7 \, _3F_2\left(\frac{5}{6},\frac{3}{2},\frac{13}{6};1,2;-\frac{9}{16} u^2 \kappa ^2\right).\label{Nu} 
\eea 
The particle spectrum dies off at large times as $u^{-1}$, so the total number indeed diverges. 

Turning to the energy, a  consistency check can be done by Hankel transforming the quantum of energy $\omega N_\omega$, and integrating over all time.  The result for the transform, $E_u =H(\omega N_\omega)/2$, is
\be E_u = \frac{\sinh \theta}{3 \sqrt{3} \pi  \kappa  u^3}-\frac{\cosh \theta}{3 \sqrt{3} \pi  u^2 \sqrt{9 \kappa ^2 u^2+16}}\,,\label{Eu}\ee
where $\theta \equiv \frac{1}{3} \sinh ^{-1}\left(\frac{3 \kappa  u}{4}\right)$. 
Eq.~(\ref{Eu}) dies off as $u^{-8/3}$ for large times, so the total 
energy is finite. The result for the total energy by integrating over all time is also analytic,  
\be E = \int_{-\infty}^{+\infty} E_u \; du = \frac{\kappa}{96}\,,\ee
which agrees with the total energy as derived by the stress tensor, Eq.~(\ref{totalE1}), and the total energy as derived by integration of the particle spectrum with respect to frequency, Eq.~(\ref{Nw_E}). 

\begin{figure}[H]
\centering
\includegraphics[width=
\columnwidth]{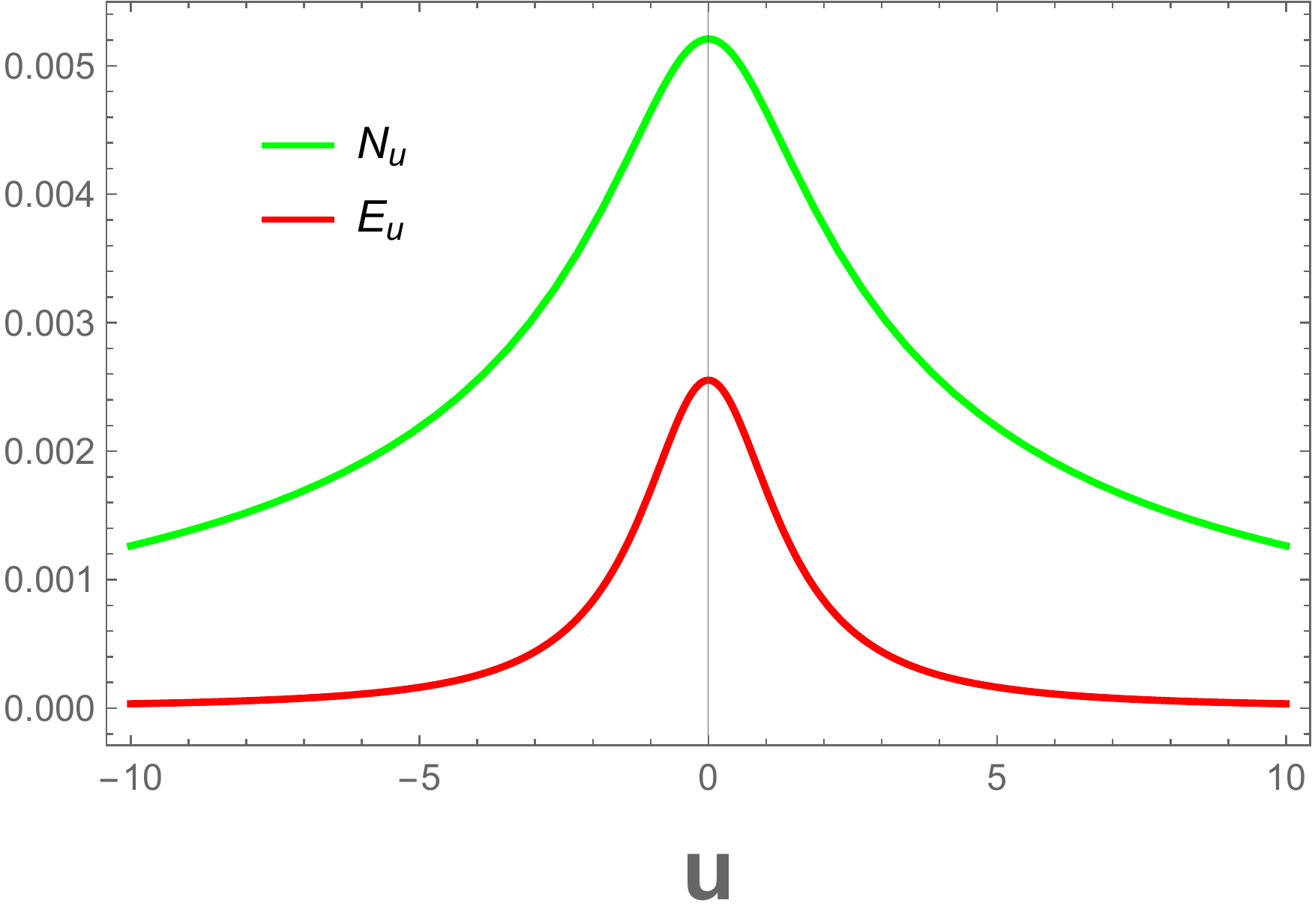}
\caption{The continuous time evolution of particle creation, Eq.~(\ref{Nu}), and time evolution of energy quanta, Eq.~(\ref{Eu}). Here $\kappa=1$ (though $N_u/\kp$ and $E_u/\kp^2$ have invariant forms as a function of $\kp u$). }\label{Fig8}
\end{figure}

As far as we know, this is the first solution for analytic time evolution of particle production from the quantum vacuum.  Notice there is no need to resort to wavepacket discreteness as the creation is continuous. Nor have we made any analytic approximations. A plot of the evolution is given in Figure \ref{Fig8}.

\section{Conclusions}

An interesting connection between the reality of the beta 
Bogoliubov coefficient, asymptotic inertia and finite energy, 
and mirror motion near the  speed of light leads to 
particle radiation by quantum vacuum that is analytic in the energy 
flux, simple in the particle spectrum -- an Airy function --  and, 
remarkably, analytic expression of the time evolution of 
particle creation. 

We evaluate the simplest allowed accelerated mirror with the 
needed conditions and derive all these physical quantities. 
The Airy mirror is asymptotically inertial, coasting at the speed of 
light; the total energy radiated is finite and simply $\kappa/96$ 
despite a soft particle divergence; the beta Bogoliubov 
coefficient is given by a real Airy-Ai function; the particle 
creation time evolution is analytic and exact. 

The mirror has no horizon, and so there is no information loss. 
The finite energy corresponds to the black hole analog case where  
evaporation ceases, related to extremal black holes, remnants, or 
complete evaporation. The asymptotic inertia is responsible for finite energy, but inertial motions that asymptotically approach the speed of light do not preserve the interpretation of entanglement entropy derived from the rapidity 
as an adequate measure of unitarity 
(see Appendix~\ref{sec:apxentropy}). 

The radiated flux exhibits regions of negative energy flux (NEF); 
these are required by unitarity for the conditions present, and 
we expand on this ``necessity of negativity'' in the Appendices, 
showing it follows directly from the asymptotically inertial (the lack of a horizon ensures information conservation\footnote{Information loss occurs from an inertial horizon \cite{Good:2020uff}.}) nature. We further connect it 
to the 1-D Schr{\"o}dinger equation and interpretation of the 
rapidity as a Lorentz transformation and wavefunction in a potential 
well defined by the acceleration properties. 

While obtaining a real, and simple, Bogoliubov coefficient is a  
significant advance, we further derive an analytic particle 
spectrum (integrating over the beta coefficient squared), 
time evolution (through a Hankel transform), and energy (further 
integrating over the spectrum times frequency). An exact analytic 
time evolution solution for particle production from 
the quantum vacuum may be unique in the literature. No discrete 
wave packetization is required (although we also show those results, 
consistent with the analytic one). 

The techniques of accelerating boundary correspondences (ABC) 
and moving mirrors continue to deliver intriguing insights into 
connections between acceleration (or surface gravity),  particle creation, and information. Furthermore, these lead to interesting 
directions for research in the properties of black holes (for 
which they serve as analogs) and quantum information, entanglement, 
and gravity.

\vspace{6pt}

\acknowledgments 

MG acknowledges funding from state-targeted program ``Center of Excellence for Fundamental and Applied Physics" (BR05236454) by the Ministry of Education and Science of the Republic of Kazakhstan, and the FY2018-SGP-1-STMM Faculty Development Competitive Research Grant No.\ 090118FD5350 at Nazarbayev University.
This work is supported in part by the Energetic Cosmos Laboratory. EL is supported in part by the U.S.\ Department of Energy, Office of Science, Office of High Energy Physics, under contract no.\ DE-AC02-05CH11231.

\appendix 

\section{Necessity of Negativity} \label{sec:apxsum} 

We emphasize that negative energy flux is a common, and 
indeed required, component of certain acceleration dynamics. 
That this follows from unitarity is discussed in 
\cite{Good:2019tnf} and references therein. Here we give 
two quick derivations. 

From Eq.~\eqref{F(v)} and the relations $f'(v)=e^{-2\eta}$ 
and $\alpha(v)= \eta'(v)\,e^\eta$, 
we can write 
\be 
24\pi F(v)=-2e^{4\eta}\left[\eta''+(\eta')^2\right]=-2e^{3\eta}\alpha'(v)\,. 
\ee 
This immediately implies 
\be 
-12\pi\int_{-\infty}^\infty dv\,e^{-3\eta}F(v)=\int_{-\infty}^\infty dv\,\frac{d\alpha}{dv}=\alpha\big|^\infty_{-\infty}\,. 
\ee 
Whenever the acceleration $\alpha$ vanishes asymptotically 
-- as it does for any asymptotically inertial dynamics -- 
(or if it is time symmetric), then the left hand side must 
be zero. Since $e^{-3\eta}$ is positive, then $F(v)$ must 
have negative regions. 

This depends only on the conditions mentioned in the 
previous paragraph and not on the specific mirror trajectory 
used in this paper. One can also see this even more directly 
in terms of proper time $\tau$: 
\be 
12\pi F(\tau)=-\alpha'(\tau)\,e^{2\eta(\tau)}\,, 
\ee 
so 
\be 
12\pi \int d\tau\,e^{-2\eta}F(\tau)= -\int d\tau\,\frac{d\alpha}{d\tau}\,. \label{eq:sum} 
\ee

\section{Zero-Energy Resonance} \label{sec:apxzero} 

The simple harmonic oscillator is the basis of many diverse 
physics areas. Here we consider a relation between particle 
radiation from an accelerated system and the oscillator equation. 
Let us adapt the usual form, $\ddot{\phi}(t) + \omega(t)\phi(t) = 0$ 
(in the time domain) or $\phi''(x) + k(x)\phi(x) = 0$ (in the space 
domain) and write it in terms of the light-cone coordinate retarded 
time $u=t-x$, 
\be 
\psi''(u)+V(u)\,\psi(u)=0\,. \label{eq:shopsi} 
\ee 
We allow the resonance frequency or spring constant to be 
spacetime dependent, and write it as $V(u)$ for reasons discussed 
below. 

We immediately have the consequence that 
\be 
\int_{-\infty}^{+\infty} du\,V(u)\,\psi(u)=-\int du\,\frac{d\psi'}{du}\,. \label{eq:sumvpsi}
\ee 
This looks quite similar to Eq.~\eqref{eq:sum}. If $\psi'$ vanishes 
at asymptotically early and late times, $|u|\to\infty$, then we 
find that for positive $\psi$ the ``potential'' $V$ must have 
negative regions. 

Let us make the analogy more concrete. If $\psi(u)=e^{-\eta}$ 
then $-\psi'(u)=\alpha(u)$, the acceleration\footnote{While $\psi'(u) = -\alpha$, it is worth pointing out that $\psi(u)$ itself is the Lorentz transformation (LT) in retarded time from un-tilded to tilded boosted frame $\tilde u = e^{-\eta}\, u$.  Here the LT acts like a wave function.}; so our constraint  
on $\psi'$ vanishing at infinity is exactly our condition in 
Appendix~\ref{sec:apxsum}, and the asymptotically 
inertial case we treat in the main text. 
Note that indeed $\psi$ is always positive. 
Now the derivatives of $\eta$, and  
hence $\psi$, are also related through the Schwarzian in 
Eq.~\eqref{F(u)} to the energy flux $F(u)$ -- which arises from 
the acceleration -- through 
\be 
V(u)\equiv 12\pi F(u) = -\frac{1}{2}\{p(u),u\}= \eta'(u)^2 - \eta''(u)\,.\label{V(u)} 
\ee 
Under these definitions, Eq.~\eqref{eq:sumvpsi} is identical 
to Eq.~\eqref{eq:sum}. Thus again we see the 
``necessity of negativity''. 

The derivation in Appendix~\ref{sec:apxsum} relied on accelerating 
system dynamics while the one here arose from the simple harmonic 
oscillator equation. The harmonic oscillator can also be related  
to the 1-D Schr{\"o}dinger equation 
\be 
-\frac{\hbar^2}{2m} \psi'' + V\psi = E \psi\,, \label{Schr1}
\ee 
for a spacetime-dependent potential where the ``spring constant'' 
\be 
k \leftrightarrow \frac{2m(E-V)}{\hbar^2}\,. 
\ee 
Absorbing the $\hbar$ and $m$ factors, and taking the zero energy 
case, we see we can rewrite the Schr{\"o}dinger equation as 
Eq.~\eqref{eq:shopsi}. Hence our $V(u)=12\pi F(u)$ does act like 
a potential and $\psi(u)$ acts like a wave function.  The moving mirror differential equation for energy flux, Eq.~(\ref{V(u)}), and the zero-energy case with absorption of a negative sign into the definition of the potential, Eq.~\eqref{Schr1}, corresponds to the physics of resonance transmission for a potential, $V(u) = V(-u)$, of a 1-D scattering threshold anomaly \cite{threshold}.

For the particular trajectory of the main text, we have the 
asymptotic condition $\psi'\to0$ but to keep the wave function 
zero at infinity we perform a 
parity flip, $x\to -x$, on the mirror trajectory $f(v)$, Eq.~(\ref{f(v)}), resulting in 
\be 
p(u) = u + \kappa^2 \frac{u^3}{3}\,.\label{p(u)} 
\ee
With $2\eta(u) = \ln p'(u)$, the rapidity $\eta(u) = \frac{1}{2} \ln (\kappa ^2 u^2+1)$, hence asymptotically $+\infty$ rather than $-\infty$ without the parity flip, i.e.\ the mirror approaches an observer located at $\mathscr{I}^+_R$ at the speed of light, instead of receding at the speed of light as is the case with Eq.~(\ref{f(v)}).  

\begin{figure}[H]
\centering
\includegraphics[width=3.5 in]{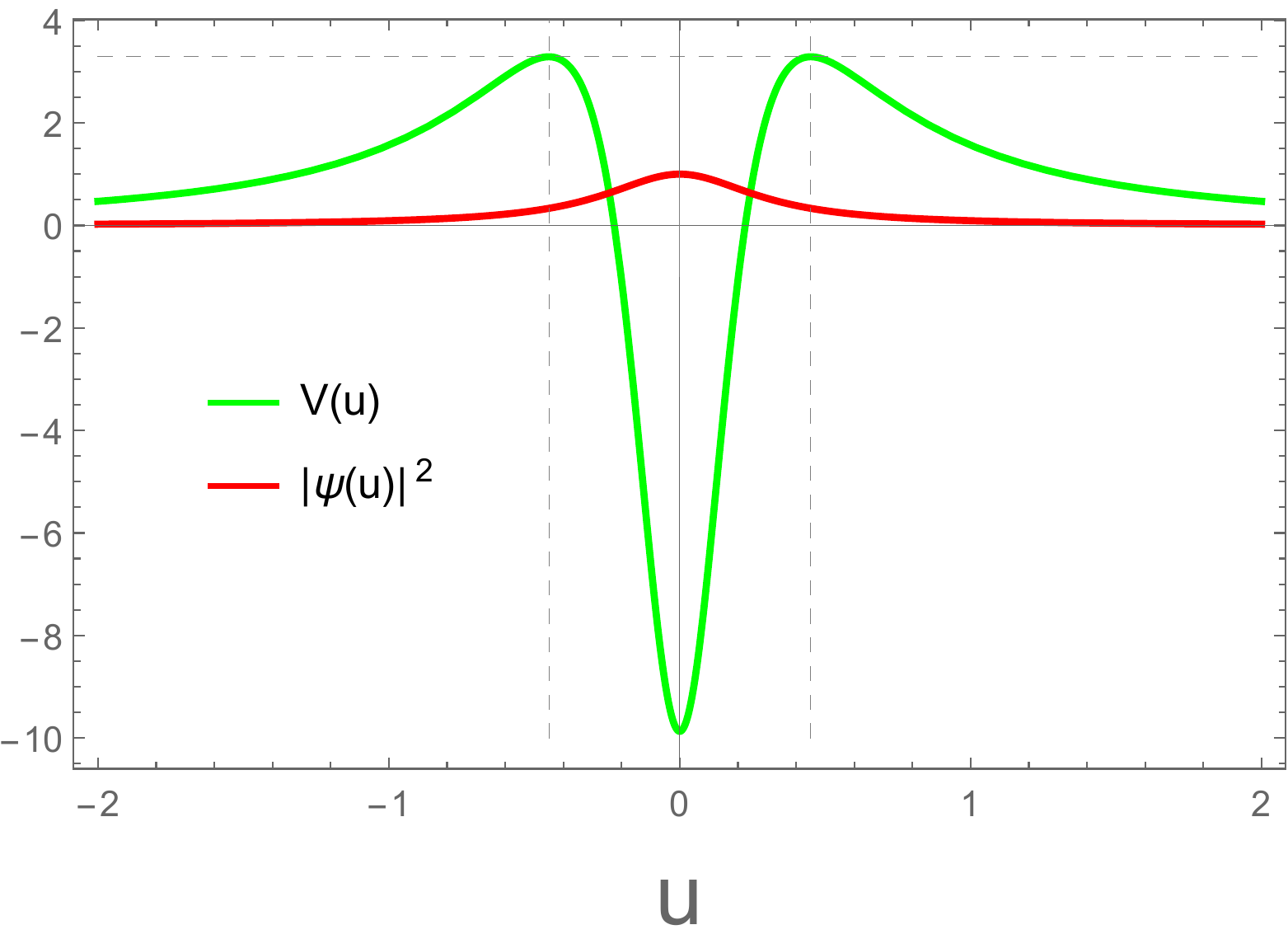}
\caption{The potential Eq.~(\ref{V(u)}) with  Eq.~(\ref{F(u)exact}), and the wave function Eq.~(\ref{psi}). $|\psi^2|$ is normalized according to Eq.~(\ref{psi2}) where $\kappa =\pi$.
The potential maxima occur at $u_m =\pm\sqrt{2}/\kappa$ with maximum value $V_m(u_m) = \kappa^2/3$; the zero crossings are at $u_0=\pm1/(\kp\sqrt{2})$. 
}\label{Fig9}
\end{figure}

The wave function form is then 
\be 
\psi(u) = \frac{1}{\sqrt{\kappa ^2 u^2+1}}\,,\qquad \psi(\pm\infty) = 0\,,\label{psi} 
\ee
plotted in Figure~\ref{Fig9}. The wave function is normalized by setting $\kappa = \pi$ so 
\be 
\int_{-\infty}^{\infty} |\psi(u)|^2 \d u = 1\,.\label{psi2} 
\ee
Plugging Eq.~(\ref{p(u)}) into the Schwarzian relation, Eq.~(\ref{F(u)}), gives 
\be 
F(u) = \frac{\kp^2 (2\kappa ^2 u^2-1)}{12 \pi  \left(\kappa ^2 u^2+1\right)^2}\,.\label{F(u)exact} 
\ee 
which is PT symmetric $u\to-u$. Phrasing this as the potential 
$V(u)=12\pi F(u)$ of the Schr{\"o}dinger equation we see in 
Figure~\ref{Fig9} how the wave function is localized within the 
potential well.

\section{Entanglement Entropy and the Speed of Light} \label{sec:apxentropy} 

Entropy diverges because rapidity does, $S=-\eta/6$. Interestingly, a divergent information measure like entanglement entropy is, at first glance, seemingly at odds with the obvious unitarity of the dynamics as seen in the Penrose diagram. However, the entanglement-rapidity formula has a subtle caveat in that it was carefully derived \cite{Myrzakul:2021bgj,Fitkevich:2020okl,Good:2015nja,Bianchi:2014qua,Chen:2017lum} assuming unitarity a priori only in the cases where entropy (rapidity) achieves a constant non-infinite value in the far future.  Since this is not the case for an asymptotic light speed moving mirror, the entropy as rapidity interpretation is not a good measure of unitarity \cite{Bianchi:2014qua} for such cases.  This example highlights the need for caution because the entanglement as rapidity approach may not hold much utility for general motions that approach the speed of light, so $\eta \to \infty$.

\bibliography{main}

\begin{thebibliography}{40}%
\makeatletter
\providecommand \@ifxundefined [1]{%
 \@ifx{#1\undefined}
}%
\providecommand \@ifnum [1]{%
 \ifnum #1\expandafter \@firstoftwo
 \else \expandafter \@secondoftwo
 \fi
}%
\providecommand \@ifx [1]{%
 \ifx #1\expandafter \@firstoftwo
 \else \expandafter \@secondoftwo
 \fi
}%
\providecommand \natexlab [1]{#1}%
\providecommand \enquote  [1]{``#1''}%
\providecommand \bibnamefont  [1]{#1}%
\providecommand \bibfnamefont [1]{#1}%
\providecommand \citenamefont [1]{#1}%
\providecommand \href@noop [0]{\@secondoftwo}%
\providecommand \href [0]{\begingroup \@sanitize@url \@href}%
\providecommand \@href[1]{\@@startlink{#1}\@@href}%
\providecommand \@@href[1]{\endgroup#1\@@endlink}%
\providecommand \@sanitize@url [0]{\catcode `\\12\catcode `\$12\catcode
  `\&12\catcode `\#12\catcode `\^12\catcode `\_12\catcode `\%12\relax}%
\providecommand \@@startlink[1]{}%
\providecommand \@@endlink[0]{}%
\providecommand \url  [0]{\begingroup\@sanitize@url \@url }%
\providecommand \@url [1]{\endgroup\@href {#1}{\urlprefix }}%
\providecommand \urlprefix  [0]{URL }%
\providecommand \Eprint [0]{\href }%
\providecommand \doibase [0]{http://dx.doi.org/}%
\providecommand \selectlanguage [0]{\@gobble}%
\providecommand \bibinfo  [0]{\@secondoftwo}%
\providecommand \bibfield  [0]{\@secondoftwo}%
\providecommand \translation [1]{[#1]}%
\providecommand \BibitemOpen [0]{}%
\providecommand \bibitemStop [0]{}%
\providecommand \bibitemNoStop [0]{.\EOS\space}%
\providecommand \EOS [0]{\spacefactor3000\relax}%
\providecommand \BibitemShut  [1]{\csname bibitem#1\endcsname}%
\let\auto@bib@innerbib\@empty
\bibitem [{\citenamefont {DeWitt}(1975)}]{DeWitt:1975ys}%
  \BibitemOpen
  \bibfield  {author} {\bibinfo {author} {\bibfnamefont {B.~S.}\ \bibnamefont
  {DeWitt}},\ }\href {\doibase 10.1016/0370-1573(75)90051-4} {\bibfield
  {journal} {\bibinfo  {journal} {Phys. Rept.}\ }\textbf {\bibinfo {volume}
  {19}},\ \bibinfo {pages} {295} (\bibinfo {year} {1975})}\BibitemShut
  {NoStop}%
\bibitem [{\citenamefont {Fulling}\ and\ \citenamefont
  {Davies}(1976)}]{Davies:1976hi}%
  \BibitemOpen
  \bibfield  {author} {\bibinfo {author} {\bibfnamefont {S.~A.}\ \bibnamefont
  {Fulling}}\ and\ \bibinfo {author} {\bibfnamefont {P.~C.~W.}\ \bibnamefont
  {Davies}},\ }\href
  {https://royalsocietypublishing.org/doi/abs/10.1098/rspa.1976.0045}
  {\bibfield  {journal} {\bibinfo  {journal} {Proceedings of the Royal Society
  of London. A. Mathematical and Physical Sciences}\ }\textbf {\bibinfo
  {volume} {348}},\ \bibinfo {pages} {393} (\bibinfo {year}
  {1976})}\BibitemShut {NoStop}%
\bibitem [{\citenamefont {Davies}\ and\ \citenamefont
  {Fulling}(1977)}]{Davies:1977yv}%
  \BibitemOpen
  \bibfield  {author} {\bibinfo {author} {\bibfnamefont {P.}~\bibnamefont
  {Davies}}\ and\ \bibinfo {author} {\bibfnamefont {S.}~\bibnamefont
  {Fulling}},\ }\href {\doibase 10.1098/rspa.1977.0130} {\bibfield  {journal}
  {\bibinfo  {journal} {Proceedings of the Royal Society of London. A.
  Mathematical and Physical Sciences}\ }\textbf {\bibinfo {volume} {A356}},\
  \bibinfo {pages} {237} (\bibinfo {year} {1977})}\BibitemShut {NoStop}%
\bibitem [{\citenamefont {Walker}\ and\ \citenamefont
  {Davies}(1982)}]{Walker_1982}%
  \BibitemOpen
  \bibfield  {author} {\bibinfo {author} {\bibfnamefont {W.~R.}\ \bibnamefont
  {Walker}}\ and\ \bibinfo {author} {\bibfnamefont {P.~C.~W.}\ \bibnamefont
  {Davies}},\ }\href {\doibase 10.1088/0305-4470/15/9/008} {\bibfield
  {journal} {\bibinfo  {journal} {Journal of Physics A: Mathematical and
  General}\ }\textbf {\bibinfo {volume} {15}},\ \bibinfo {pages} {L477}
  (\bibinfo {year} {1982})}\BibitemShut {NoStop}%
\bibitem [{\citenamefont {Good}\ \emph {et~al.}(2016)\citenamefont {Good},
  \citenamefont {Anderson},\ and\ \citenamefont {Evans}}]{Good:2016oey}%
  \BibitemOpen
  \bibfield  {author} {\bibinfo {author} {\bibfnamefont {M.~R.~R.}\
  \bibnamefont {Good}}, \bibinfo {author} {\bibfnamefont {P.~R.}\ \bibnamefont
  {Anderson}}, \ and\ \bibinfo {author} {\bibfnamefont {C.~R.}\ \bibnamefont
  {Evans}},\ }\href {\doibase 10.1103/PhysRevD.94.065010} {\bibfield  {journal}
  {\bibinfo  {journal} {Phys. Rev. D}\ }\textbf {\bibinfo {volume} {94}},\
  \bibinfo {pages} {065010} (\bibinfo {year} {2016})},\ \Eprint
  {http://arxiv.org/abs/1605.06635} {arXiv:1605.06635 [gr-qc]} \BibitemShut
  {NoStop}%
\bibitem [{\citenamefont {Good}\ and\ \citenamefont
  {Ong}(2020)}]{good2020particle}%
  \BibitemOpen
  \bibfield  {author} {\bibinfo {author} {\bibfnamefont {M.~R.~R.}\
  \bibnamefont {Good}}\ and\ \bibinfo {author} {\bibfnamefont {Y.~C.}\
  \bibnamefont {Ong}},\ }\href {\doibase 10.1140/epjc/s10052-020-08761-7}
  {\bibfield  {journal} {\bibinfo  {journal} {Eur. Phys. J. C}\ }\textbf
  {\bibinfo {volume} {80}},\ \bibinfo {pages} {1169} (\bibinfo {year}
  {2020})},\ \Eprint {http://arxiv.org/abs/2004.03916} {arXiv:2004.03916
  [gr-qc]} \BibitemShut {NoStop}%
\bibitem [{\citenamefont {Good}\ \emph
  {et~al.}(2020{\natexlab{a}})\citenamefont {Good}, \citenamefont {Foo},\ and\
  \citenamefont {Linder}}]{Good:2020fjz}%
  \BibitemOpen
  \bibfield  {author} {\bibinfo {author} {\bibfnamefont {M.~R.}\ \bibnamefont
  {Good}}, \bibinfo {author} {\bibfnamefont {J.}~\bibnamefont {Foo}}, \ and\
  \bibinfo {author} {\bibfnamefont {E.~V.}\ \bibnamefont {Linder}},\
  }\href@noop {} {\  (\bibinfo {year} {2020}{\natexlab{a}})},\ \Eprint
  {http://arxiv.org/abs/2006.01349} {arXiv:2006.01349 [gr-qc]} \BibitemShut
  {NoStop}%
\bibitem [{\citenamefont {Good}\ \emph
  {et~al.}(2020{\natexlab{b}})\citenamefont {Good}, \citenamefont {Zhakenuly},\
  and\ \citenamefont {Linder}}]{Good:2020byh}%
  \BibitemOpen
  \bibfield  {author} {\bibinfo {author} {\bibfnamefont {M.~R.~R.}\
  \bibnamefont {Good}}, \bibinfo {author} {\bibfnamefont {A.}~\bibnamefont
  {Zhakenuly}}, \ and\ \bibinfo {author} {\bibfnamefont {E.~V.}\ \bibnamefont
  {Linder}},\ }\href {\doibase 10.1103/PhysRevD.102.045020} {\bibfield
  {journal} {\bibinfo  {journal} {Phys. Rev. D}\ }\textbf {\bibinfo {volume}
  {102}},\ \bibinfo {pages} {045020} (\bibinfo {year} {2020}{\natexlab{b}})},\
  \Eprint {http://arxiv.org/abs/2005.03850} {arXiv:2005.03850 [gr-qc]}
  \BibitemShut {NoStop}%
\bibitem [{\citenamefont {Liberati}\ \emph {et~al.}(2000)\citenamefont
  {Liberati}, \citenamefont {Rothman},\ and\ \citenamefont
  {Sonego}}]{Liberati:2000sq}%
  \BibitemOpen
  \bibfield  {author} {\bibinfo {author} {\bibfnamefont {S.}~\bibnamefont
  {Liberati}}, \bibinfo {author} {\bibfnamefont {T.}~\bibnamefont {Rothman}}, \
  and\ \bibinfo {author} {\bibfnamefont {S.}~\bibnamefont {Sonego}},\ }\href
  {\doibase 10.1103/PhysRevD.62.024005} {\bibfield  {journal} {\bibinfo
  {journal} {Phys. Rev. D}\ }\textbf {\bibinfo {volume} {62}},\ \bibinfo
  {pages} {024005} (\bibinfo {year} {2000})},\ \Eprint
  {http://arxiv.org/abs/gr-qc/0002019} {arXiv:gr-qc/0002019} \BibitemShut
  {NoStop}%
\bibitem [{\citenamefont {Good}(2020)}]{good2020extreme}%
  \BibitemOpen
  \bibfield  {author} {\bibinfo {author} {\bibfnamefont {M.~R.~R.}\
  \bibnamefont {Good}},\ }\href {\doibase 10.1103/PhysRevD.101.104050}
  {\bibfield  {journal} {\bibinfo  {journal} {Phys. Rev. D}\ }\textbf {\bibinfo
  {volume} {101}},\ \bibinfo {pages} {104050} (\bibinfo {year} {2020})},\
  \Eprint {http://arxiv.org/abs/2003.07016} {arXiv:2003.07016 [gr-qc]}
  \BibitemShut {NoStop}%
\bibitem [{\citenamefont {Rothman}(2000)}]{Rothman:2000mm}%
  \BibitemOpen
  \bibfield  {author} {\bibinfo {author} {\bibfnamefont {T.}~\bibnamefont
  {Rothman}},\ }\href {\doibase 10.1016/S0375-9601(00)00515-6} {\bibfield
  {journal} {\bibinfo  {journal} {Phys. Lett. A}\ }\textbf {\bibinfo {volume}
  {273}},\ \bibinfo {pages} {303} (\bibinfo {year} {2000})},\ \Eprint
  {http://arxiv.org/abs/gr-qc/0006036} {arXiv:gr-qc/0006036} \BibitemShut
  {NoStop}%
\bibitem [{\citenamefont {Foo}\ and\ \citenamefont {Good}(2021)}]{Foo:2020bmv}%
  \BibitemOpen
  \bibfield  {author} {\bibinfo {author} {\bibfnamefont {J.}~\bibnamefont
  {Foo}}\ and\ \bibinfo {author} {\bibfnamefont {M.~R.~R.}\ \bibnamefont
  {Good}},\ }\href {\doibase 10.1088/1475-7516/2021/01/019} {\bibfield
  {journal} {\bibinfo  {journal} {JCAP}\ }\textbf {\bibinfo {volume} {01}},\
  \bibinfo {pages} {019} (\bibinfo {year} {2021})},\ \Eprint
  {http://arxiv.org/abs/2006.09681} {arXiv:2006.09681 [gr-qc]} \BibitemShut
  {NoStop}%
\bibitem [{\citenamefont {Good}\ \emph {et~al.}(2017)\citenamefont {Good},
  \citenamefont {Yelshibekov},\ and\ \citenamefont {Ong}}]{Good:2016atu}%
  \BibitemOpen
  \bibfield  {author} {\bibinfo {author} {\bibfnamefont {M.~R.~R.}\
  \bibnamefont {Good}}, \bibinfo {author} {\bibfnamefont {K.}~\bibnamefont
  {Yelshibekov}}, \ and\ \bibinfo {author} {\bibfnamefont {Y.~C.}\ \bibnamefont
  {Ong}},\ }\href {\doibase 10.1007/JHEP03(2017)013} {\bibfield  {journal}
  {\bibinfo  {journal} {JHEP}\ }\textbf {\bibinfo {volume} {03}},\ \bibinfo
  {pages} {013} (\bibinfo {year} {2017})},\ \Eprint
  {http://arxiv.org/abs/1611.00809} {arXiv:1611.00809 [gr-qc]} \BibitemShut
  {NoStop}%
\bibitem [{\citenamefont {Good}\ \emph {et~al.}(2019)\citenamefont {Good},
  \citenamefont {Ong}, \citenamefont {Myrzakul},\ and\ \citenamefont
  {Yelshibekov}}]{Good:2018ell}%
  \BibitemOpen
  \bibfield  {author} {\bibinfo {author} {\bibfnamefont {M.~R.}\ \bibnamefont
  {Good}}, \bibinfo {author} {\bibfnamefont {Y.~C.}\ \bibnamefont {Ong}},
  \bibinfo {author} {\bibfnamefont {A.}~\bibnamefont {Myrzakul}}, \ and\
  \bibinfo {author} {\bibfnamefont {K.}~\bibnamefont {Yelshibekov}},\ }\href
  {\doibase 10.1007/s10714-019-2575-5} {\bibfield  {journal} {\bibinfo
  {journal} {Gen. Rel. Grav.}\ }\textbf {\bibinfo {volume} {51}},\ \bibinfo
  {pages} {92} (\bibinfo {year} {2019})},\ \Eprint
  {http://arxiv.org/abs/1801.08020} {arXiv:1801.08020 [gr-qc]} \BibitemShut
  {NoStop}%
\bibitem [{\citenamefont {Good}(2018)}]{Good:2018zmx}%
  \BibitemOpen
  \bibfield  {author} {\bibinfo {author} {\bibfnamefont {M.~R.}\ \bibnamefont
  {Good}},\ }\href {\doibase 10.3390/universe4110122} {\bibfield  {journal}
  {\bibinfo  {journal} {Universe}\ }\textbf {\bibinfo {volume} {4}},\ \bibinfo
  {pages} {122} (\bibinfo {year} {2018})}\BibitemShut {NoStop}%
\bibitem [{\citenamefont {Myrzakul}\ and\ \citenamefont
  {Good}(2018)}]{Myrzakul:2018bhy}%
  \BibitemOpen
  \bibfield  {author} {\bibinfo {author} {\bibfnamefont {A.}~\bibnamefont
  {Myrzakul}}\ and\ \bibinfo {author} {\bibfnamefont {M.~R.}\ \bibnamefont
  {Good}},\ }in\ \href@noop {} {\emph {\bibinfo {booktitle} {{15th Marcel
  Grossmann Meeting on Recent Developments in Theoretical and Experimental
  General Relativity, Astrophysics, and Relativistic Field Theories}}}}\
  (\bibinfo {year} {2018})\ \Eprint {http://arxiv.org/abs/1807.10627}
  {arXiv:1807.10627 [gr-qc]} \BibitemShut {NoStop}%
\bibitem [{\citenamefont {Good}\ and\ \citenamefont
  {Ong}(2015)}]{Good:2015nja}%
  \BibitemOpen
  \bibfield  {author} {\bibinfo {author} {\bibfnamefont {M.~R.~R.}\
  \bibnamefont {Good}}\ and\ \bibinfo {author} {\bibfnamefont {Y.~C.}\
  \bibnamefont {Ong}},\ }\href {\doibase 10.1007/JHEP07(2015)145} {\bibfield
  {journal} {\bibinfo  {journal} {JHEP}\ }\textbf {\bibinfo {volume} {07}},\
  \bibinfo {pages} {145} (\bibinfo {year} {2015})},\ \Eprint
  {http://arxiv.org/abs/1506.08072} {arXiv:1506.08072 [gr-qc]} \BibitemShut
  {NoStop}%
\bibitem [{\citenamefont {Good}(2017)}]{Good:2016yht}%
  \BibitemOpen
  \bibfield  {author} {\bibinfo {author} {\bibfnamefont {M.~R.~R.}\
  \bibnamefont {Good}},\ }\href {\doibase
  https://dx.doi.org/10.1142/9789813207431_0014} {\emph {\bibinfo {title}
  {{Reflecting at the Speed of Light}}}}\ (\bibinfo  {publisher} {World
  Scientific},\ \bibinfo {address} {Singapore},\ \bibinfo {year} {2017})\
  \Eprint {http://arxiv.org/abs/1612.02459} {arXiv:1612.02459 [gr-qc]}
  \BibitemShut {NoStop}%
\bibitem [{\citenamefont {Good}\ \emph
  {et~al.}(2020{\natexlab{c}})\citenamefont {Good}, \citenamefont {Linder},\
  and\ \citenamefont {Wilczek}}]{Good:2019tnf}%
  \BibitemOpen
  \bibfield  {author} {\bibinfo {author} {\bibfnamefont {M.~R.}\ \bibnamefont
  {Good}}, \bibinfo {author} {\bibfnamefont {E.~V.}\ \bibnamefont {Linder}}, \
  and\ \bibinfo {author} {\bibfnamefont {F.}~\bibnamefont {Wilczek}},\ }\href
  {\doibase 10.1103/PhysRevD.101.025012} {\bibfield  {journal} {\bibinfo
  {journal} {Phys. Rev. D}\ }\textbf {\bibinfo {volume} {101}},\ \bibinfo
  {pages} {025012} (\bibinfo {year} {2020}{\natexlab{c}})},\ \Eprint
  {http://arxiv.org/abs/1909.01129} {arXiv:1909.01129 [gr-qc]} \BibitemShut
  {NoStop}%
\bibitem [{\citenamefont {Good}\ \emph
  {et~al.}(2020{\natexlab{d}})\citenamefont {Good}, \citenamefont {Linder},\
  and\ \citenamefont {Wilczek}}]{GoodMPLA}%
  \BibitemOpen
  \bibfield  {author} {\bibinfo {author} {\bibfnamefont {M.~R.~R.}\
  \bibnamefont {Good}}, \bibinfo {author} {\bibfnamefont {E.~V.}\ \bibnamefont
  {Linder}}, \ and\ \bibinfo {author} {\bibfnamefont {F.}~\bibnamefont
  {Wilczek}},\ }\href {\doibase 10.1142/S0217732320400064} {\bibfield
  {journal} {\bibinfo  {journal} {Modern Physics Letters A}\ }\textbf {\bibinfo
  {volume} {35}},\ \bibinfo {pages} {2040006} (\bibinfo {year}
  {2020}{\natexlab{d}})}\BibitemShut {NoStop}%
\bibitem [{\citenamefont {Good}\ and\ \citenamefont
  {Linder}(2017)}]{Good:2017kjr}%
  \BibitemOpen
  \bibfield  {author} {\bibinfo {author} {\bibfnamefont {M.~R.~R.}\
  \bibnamefont {Good}}\ and\ \bibinfo {author} {\bibfnamefont {E.~V.}\
  \bibnamefont {Linder}},\ }\href {\doibase 10.1103/PhysRevD.96.125010}
  {\bibfield  {journal} {\bibinfo  {journal} {Phys. Rev. D}\ }\textbf {\bibinfo
  {volume} {96}},\ \bibinfo {pages} {125010} (\bibinfo {year} {2017})},\
  \Eprint {http://arxiv.org/abs/1707.03670} {arXiv:1707.03670 [gr-qc]}
  \BibitemShut {NoStop}%
\bibitem [{\citenamefont {Good}\ \emph {et~al.}(2013)\citenamefont {Good},
  \citenamefont {Anderson},\ and\ \citenamefont {Evans}}]{good2013time}%
  \BibitemOpen
  \bibfield  {author} {\bibinfo {author} {\bibfnamefont {M.~R.~R.}\
  \bibnamefont {Good}}, \bibinfo {author} {\bibfnamefont {P.~R.}\ \bibnamefont
  {Anderson}}, \ and\ \bibinfo {author} {\bibfnamefont {C.~R.}\ \bibnamefont
  {Evans}},\ }\href {\doibase 10.1103/PhysRevD.88.025023} {\bibfield  {journal}
  {\bibinfo  {journal} {Phys. Rev. D}\ }\textbf {\bibinfo {volume} {88}},\
  \bibinfo {pages} {025023} (\bibinfo {year} {2013})},\ \Eprint
  {http://arxiv.org/abs/1303.6756} {arXiv:1303.6756 [gr-qc]} \BibitemShut
  {NoStop}%
\bibitem [{\citenamefont {Good}\ and\ \citenamefont
  {Linder}(2018)}]{Good:2017ddq}%
  \BibitemOpen
  \bibfield  {author} {\bibinfo {author} {\bibfnamefont {M.~R.}\ \bibnamefont
  {Good}}\ and\ \bibinfo {author} {\bibfnamefont {E.~V.}\ \bibnamefont
  {Linder}},\ }\href {\doibase 10.1103/PhysRevD.97.065006} {\bibfield
  {journal} {\bibinfo  {journal} {Phys. Rev. D}\ }\textbf {\bibinfo {volume}
  {97}},\ \bibinfo {pages} {065006} (\bibinfo {year} {2018})},\ \Eprint
  {http://arxiv.org/abs/1711.09922} {arXiv:1711.09922 [gr-qc]} \BibitemShut
  {NoStop}%
\bibitem [{\citenamefont {Good}\ and\ \citenamefont
  {Linder}(2019)}]{Good:2018aer}%
  \BibitemOpen
  \bibfield  {author} {\bibinfo {author} {\bibfnamefont {M.~R.}\ \bibnamefont
  {Good}}\ and\ \bibinfo {author} {\bibfnamefont {E.~V.}\ \bibnamefont
  {Linder}},\ }\href {\doibase 10.1103/PhysRevD.99.025009} {\bibfield
  {journal} {\bibinfo  {journal} {Phys. Rev. D}\ }\textbf {\bibinfo {volume}
  {99}},\ \bibinfo {pages} {025009} (\bibinfo {year} {2019})},\ \Eprint
  {http://arxiv.org/abs/1807.08632} {arXiv:1807.08632 [gr-qc]} \BibitemShut
  {NoStop}%
\bibitem [{\citenamefont {S{\o}rensen}(2012)}]{exact}%
  \BibitemOpen
  \bibfield  {author} {\bibinfo {author} {\bibfnamefont {O.}~\bibnamefont
  {S{\o}rensen}},\ }\emph {\bibinfo {title} {Exact treatment of interacting
  bosons in rotating systems and lattices}},\ \href@noop {} {Ph.D. thesis}
  (\bibinfo {year} {2012})\BibitemShut {NoStop}%
\bibitem [{\citenamefont {Birrell}\ and\ \citenamefont
  {Davies}(1984)}]{Birrell:1982ix}%
  \BibitemOpen
  \bibfield  {author} {\bibinfo {author} {\bibfnamefont {N.}~\bibnamefont
  {Birrell}}\ and\ \bibinfo {author} {\bibfnamefont {P.}~\bibnamefont
  {Davies}},\ }\href {\doibase 10.1017/CBO9780511622632} {\emph {\bibinfo
  {title} {{Quantum Fields in Curved Space}}}},\ Cambridge Monographs on
  Mathematical Physics\ (\bibinfo  {publisher} {Cambridge Univ. Press},\
  \bibinfo {address} {Cambridge, UK},\ \bibinfo {year} {1984})\BibitemShut
  {NoStop}%
\bibitem [{\citenamefont {Carlitz}\ and\ \citenamefont
  {Willey}(1987)}]{carlitz1987reflections}%
  \BibitemOpen
  \bibfield  {author} {\bibinfo {author} {\bibfnamefont {R.~D.}\ \bibnamefont
  {Carlitz}}\ and\ \bibinfo {author} {\bibfnamefont {R.~S.}\ \bibnamefont
  {Willey}},\ }\href {\doibase 10.1103/PhysRevD.36.2327} {\bibfield  {journal}
  {\bibinfo  {journal} {Phys. Rev. D}\ }\textbf {\bibinfo {volume} {36}},\
  \bibinfo {pages} {2327} (\bibinfo {year} {1987})}\BibitemShut {NoStop}%
\bibitem [{\citenamefont {Walker}(1985)}]{walker1985particle}%
  \BibitemOpen
  \bibfield  {author} {\bibinfo {author} {\bibfnamefont {W.~R.}\ \bibnamefont
  {Walker}},\ }\href {\doibase 10.1103/PhysRevD.31.767} {\bibfield  {journal}
  {\bibinfo  {journal} {Phys. Rev. D}\ }\textbf {\bibinfo {volume} {31}},\
  \bibinfo {pages} {767} (\bibinfo {year} {1985})}\BibitemShut {NoStop}%
\bibitem [{\citenamefont {Ford}\ and\ \citenamefont
  {Roman}(1999)}]{Ford:1999qv}%
  \BibitemOpen
  \bibfield  {author} {\bibinfo {author} {\bibfnamefont {L.}~\bibnamefont
  {Ford}}\ and\ \bibinfo {author} {\bibfnamefont {T.~A.}\ \bibnamefont
  {Roman}},\ }\href {\doibase 10.1103/PhysRevD.60.104018} {\bibfield  {journal}
  {\bibinfo  {journal} {Phys. Rev. D}\ }\textbf {\bibinfo {volume} {60}},\
  \bibinfo {pages} {104018} (\bibinfo {year} {1999})},\ \Eprint
  {http://arxiv.org/abs/gr-qc/9901074} {arXiv:gr-qc/9901074} \BibitemShut
  {NoStop}%
\bibitem [{\citenamefont {Ford}\ and\ \citenamefont
  {Roman}(1995)}]{Ford:1994bj}%
  \BibitemOpen
  \bibfield  {author} {\bibinfo {author} {\bibfnamefont {L.}~\bibnamefont
  {Ford}}\ and\ \bibinfo {author} {\bibfnamefont {T.~A.}\ \bibnamefont
  {Roman}},\ }\href {\doibase 10.1103/PhysRevD.51.4277} {\bibfield  {journal}
  {\bibinfo  {journal} {Phys. Rev. D}\ }\textbf {\bibinfo {volume} {51}},\
  \bibinfo {pages} {4277} (\bibinfo {year} {1995})},\ \Eprint
  {http://arxiv.org/abs/gr-qc/9410043} {arXiv:gr-qc/9410043} \BibitemShut
  {NoStop}%
\bibitem [{\citenamefont {Nikishov}(2003)}]{Nikishov:2002ez}%
  \BibitemOpen
  \bibfield  {author} {\bibinfo {author} {\bibfnamefont {A.~I.}\ \bibnamefont
  {Nikishov}},\ }\href {\doibase 10.1134/1.1560392} {\bibfield  {journal}
  {\bibinfo  {journal} {J. Exp. Theor. Phys.}\ }\textbf {\bibinfo {volume}
  {96}},\ \bibinfo {pages} {180} (\bibinfo {year} {2003})},\ \Eprint
  {http://arxiv.org/abs/hep-th/0207085} {arXiv:hep-th/0207085} \BibitemShut
  {NoStop}%
\bibitem [{\citenamefont {Hawking}(1975)}]{Hawking:1974sw}%
  \BibitemOpen
  \bibfield  {author} {\bibinfo {author} {\bibfnamefont {S.}~\bibnamefont
  {Hawking}},\ }\href {\doibase 10.1007/BF02345020} {\bibfield  {journal}
  {\bibinfo  {journal} {Commun. Math. Phys.}\ }\textbf {\bibinfo {volume}
  {43}},\ \bibinfo {pages} {199} (\bibinfo {year} {1975})}\BibitemShut
  {NoStop}%
\bibitem [{\citenamefont {Good}(2013)}]{Good:2012cp}%
  \BibitemOpen
  \bibfield  {author} {\bibinfo {author} {\bibfnamefont {M.~R.}\ \bibnamefont
  {Good}},\ }\href {\doibase 10.1142/S0217751X13500085} {\bibfield  {journal}
  {\bibinfo  {journal} {Int. J. Mod. Phys. A}\ }\textbf {\bibinfo {volume}
  {28}},\ \bibinfo {pages} {1350008} (\bibinfo {year} {2013})},\ \Eprint
  {http://arxiv.org/abs/1205.0881} {arXiv:1205.0881 [gr-qc]} \BibitemShut
  {NoStop}%
\bibitem [{\citenamefont {Fabbri}\ and\ \citenamefont
  {Navarro-Salas}(2005)}]{Fabbri}%
  \BibitemOpen
  \bibfield  {author} {\bibinfo {author} {\bibfnamefont {A.}~\bibnamefont
  {Fabbri}}\ and\ \bibinfo {author} {\bibfnamefont {J.}~\bibnamefont
  {Navarro-Salas}},\ }\href
  {https://www.worldscientific.com/doi/abs/10.1142/p378} {\emph {\bibinfo
  {title} {Modeling Black Hole Evaporation}}}\ (\bibinfo  {publisher} {Imperial
  College Press},\ \bibinfo {year} {2005})\BibitemShut {NoStop}%
\bibitem [{\citenamefont {Good}\ and\ \citenamefont
  {Abdikamalov}(2020)}]{Good:2020uff}%
  \BibitemOpen
  \bibfield  {author} {\bibinfo {author} {\bibfnamefont {M.}~\bibnamefont
  {Good}}\ and\ \bibinfo {author} {\bibfnamefont {E.}~\bibnamefont
  {Abdikamalov}},\ }\href {\doibase 10.3390/universe6090131} {\bibfield
  {journal} {\bibinfo  {journal} {Universe}\ }\textbf {\bibinfo {volume} {6}},\
  \bibinfo {pages} {131} (\bibinfo {year} {2020})},\ \Eprint
  {http://arxiv.org/abs/2008.08776} {arXiv:2008.08776 [gr-qc]} \BibitemShut
  {NoStop}%
\bibitem [{\citenamefont {Senn}(1988)}]{threshold}%
  \BibitemOpen
  \bibfield  {author} {\bibinfo {author} {\bibfnamefont {P.}~\bibnamefont
  {Senn}},\ }\href {\doibase 10.1119/1.15359} {\bibfield  {journal} {\bibinfo
  {journal} {American Journal of Physics}\ }\textbf {\bibinfo {volume} {56}},\
  \bibinfo {pages} {916} (\bibinfo {year} {1988})}\BibitemShut {NoStop}%
\bibitem [{\citenamefont {Myrzakul}\ \emph {et~al.}(2021)\citenamefont
  {Myrzakul}, \citenamefont {Xiong},\ and\ \citenamefont
  {Good}}]{Myrzakul:2021bgj}%
  \BibitemOpen
  \bibfield  {author} {\bibinfo {author} {\bibfnamefont {A.}~\bibnamefont
  {Myrzakul}}, \bibinfo {author} {\bibfnamefont {C.}~\bibnamefont {Xiong}}, \
  and\ \bibinfo {author} {\bibfnamefont {M.~R.~R.}\ \bibnamefont {Good}},\
  }\href@noop {} {\  (\bibinfo {year} {2021})},\ \Eprint
  {http://arxiv.org/abs/2101.08139} {arXiv:2101.08139 [gr-qc]} \BibitemShut
  {NoStop}%
\bibitem [{\citenamefont {Fitkevich}\ \emph {et~al.}(2020)\citenamefont
  {Fitkevich}, \citenamefont {Levkov},\ and\ \citenamefont
  {Zenkevich}}]{Fitkevich:2020okl}%
  \BibitemOpen
  \bibfield  {author} {\bibinfo {author} {\bibfnamefont {M.}~\bibnamefont
  {Fitkevich}}, \bibinfo {author} {\bibfnamefont {D.}~\bibnamefont {Levkov}}, \
  and\ \bibinfo {author} {\bibfnamefont {Y.}~\bibnamefont {Zenkevich}},\ }\href
  {\doibase 10.1007/JHEP06(2020)184} {\bibfield  {journal} {\bibinfo  {journal}
  {JHEP}\ }\textbf {\bibinfo {volume} {20}},\ \bibinfo {pages} {184} (\bibinfo
  {year} {2020})},\ \Eprint {http://arxiv.org/abs/2004.13745} {arXiv:2004.13745
  [hep-th]} \BibitemShut {NoStop}%
\bibitem [{\citenamefont {Bianchi}\ and\ \citenamefont
  {Smerlak}(2014)}]{Bianchi:2014qua}%
  \BibitemOpen
  \bibfield  {author} {\bibinfo {author} {\bibfnamefont {E.}~\bibnamefont
  {Bianchi}}\ and\ \bibinfo {author} {\bibfnamefont {M.}~\bibnamefont
  {Smerlak}},\ }\href {\doibase 10.1103/PhysRevD.90.041904} {\bibfield
  {journal} {\bibinfo  {journal} {Phys. Rev. D}\ }\textbf {\bibinfo {volume}
  {90}},\ \bibinfo {pages} {041904} (\bibinfo {year} {2014})},\ \Eprint
  {http://arxiv.org/abs/1404.0602} {arXiv:1404.0602 [gr-qc]} \BibitemShut
  {NoStop}%
\bibitem [{\citenamefont {Chen}\ and\ \citenamefont
  {Yeom}(2017)}]{Chen:2017lum}%
  \BibitemOpen
  \bibfield  {author} {\bibinfo {author} {\bibfnamefont {P.}~\bibnamefont
  {Chen}}\ and\ \bibinfo {author} {\bibfnamefont {D.-h.}\ \bibnamefont
  {Yeom}},\ }\href {\doibase 10.1103/PhysRevD.96.025016} {\bibfield  {journal}
  {\bibinfo  {journal} {Phys. Rev. D}\ }\textbf {\bibinfo {volume} {96}},\
  \bibinfo {pages} {025016} (\bibinfo {year} {2017})},\ \Eprint
  {http://arxiv.org/abs/1704.08613} {arXiv:1704.08613 [hep-th]} \BibitemShut
  {NoStop}%
\end{thebibliography}%

\end{document}